\documentclass[11pt,twoside]{article}
\usepackage{amsmath}
\usepackage{amssymb}
\usepackage{amscd}
\def\EndBox#1{
	\hskip0.1em\hfill\null\ \null\nobreak\hfill\kern3pt
		\hbox{$\scriptstyle #1$} \smallbreak}
\def\qed{\EndBox{\square}}
\newtheorem{definition}{Definition}[section]
\newtheorem{proposition}{Proposition}[section]
\newtheorem{theorem}{Theorem}[section]
\newcommand{\proof}{{\sc proof:~}}
\newcommand{\remark}{\smallbreak\noindent{\bf Remark.}}
\renewcommand{\a}{\alpha}

\renewcommand{\b}{\beta}
\newcommand{\g}{\gamma}
\newcommand{\G}{\Gamma}
\renewcommand{\d}{\delta}
\newcommand{\ep}{\epsilon}
\newcommand{\z}{{\zeta}}
\renewcommand{\th}{\theta}

\renewcommand{\l}{\lambda}
\newcommand{\m}{\mu}

\renewcommand{\r}{\rho}
\newcommand{\s}{\sigma}
\renewcommand{\t}{\tau}

\newcommand{\om}{{\omega}}
\newcommand{\Om}{{\Omega}}
\newcommand{\Cs}{{\rlap{\lower3pt\hbox{\textnormal{\LARGE \char'040}}}{\Gamma}}{}}
\newcommand{\de}{\partial}
\newcommand{\Eo}{{\scriptstyle{\mathrm{E}}}}
\newcommand{\detg}{{{\scriptstyle|}g{\scriptstyle|}}}
\newcommand{\rdg}{{\textstyle\sqrt{{\scriptstyle|}g{\scriptstyle|}}\,}}
\newcommand{\rrdg}{{\scriptstyle\sqrt{{\sst|}g{\sst|}}}}
\newcommand{\oh}{\tfrac{1}{2}}
\newcommand{\ih}{\tfrac{\iO}{2}}
\newcommand{\oq}{\tfrac{1}{4}}

\newcommand{\dde}[2]{\frac{\partial #1}{\partial #2}}
\newcommand{\cj}[1]{\overline{#1}}
\renewcommand{\.}{{\scriptstyle\boldsymbol{\dot{}}}}
\newcommand{\sbot}{{\scriptscriptstyle\bot}}
\newcommand{\sbo}{{\!\sbot}}
\newcommand{\spar}{{\scriptscriptstyle\|}}
\newcommand{\bz}{{\bar z}}
\newcommand{\E}{{\boldsymbol{E}}}
\newcommand{\F}{{\boldsymbol{F}}}

\newcommand{\Gb}{{\boldsymbol{G}}}

\newcommand{\GA}{{\boldsymbol{\Gamma}}}
\newcommand{\GAG}{{\boldsymbol{\Gamma}}_{{\!}_\Gb}}
\newcommand{\LA}{{\boldsymbol{\Lambda}}{}}
\newcommand{\M}{{\boldsymbol{M}}}
\newcommand{\Mm}{{\scriptscriptstyle{\boldsymbol{M}}}}
\newcommand{\N}{{\boldsymbol{N}}}
\renewcommand{\P}{{\boldsymbol{P}}}
\newcommand{\Pm}{\P_{\!\!m}}
\newcommand{\T}{{\boldsymbol{T}}}
\newcommand{\Tt}{{\scriptscriptstyle{\boldsymbol{T}}}}
\newcommand{\W}{{\boldsymbol{W}}}
\newcommand{\Wc}{\cj{\W}}

\newcommand{\X}{{\boldsymbol{X}}}
\newcommand{\Xx}{{\scriptscriptstyle{\boldsymbol{X}}}}
\newcommand{\Y}{{\boldsymbol{Y}}}
\newcommand{\Z}{{\boldsymbol{Z}}}
\newcommand{\Zc}{\cj{\Z}}
\newcommand{\Za}{\cj{\Z}{}^*}
\newcommand{\Lie}{\mathfrak{L}}
\newcommand{\CC}{{\mathbb{C}}}
\newcommand{\NN}{{\mathbb{N}}}
\newcommand{\RR}{{\mathbb{R}}}
\newcommand{\VV}{{\mathbb{V}}}
\newcommand{\ZZ}{{\mathbb{Z}}}
\newcommand{\RRr}{{\scriptscriptstyle{\mathbb{R}}}}
\newcommand{\Fcal}{{\mathcal{F}}}
\newcommand{\Ical}{{\mathcal{I}}}
\newcommand{\Jcal}{{\mathcal{J}}}
\newcommand{\Kcal}{{\mathcal{K}}}
\newcommand{\Lcal}{{\mathcal{L}}}
\newcommand{\Mcal}{{\mathcal{M}}}
\newcommand{\Ncal}{{\mathcal{N}}}
\newcommand{\Pcal}{{\mathcal{P}}}
\newcommand{\Rcal}{{\mathcal{R}}}
\newcommand{\lfr}{\mathfrak{l}}
\newcommand{\CCal}{{\boldsymbol{\mathcal{C}}}}
\newcommand{\DC}{{\boldsymbol{\mathcal{D}}}}
\newcommand{\EC}{{\boldsymbol{\mathcal{E}}}}
\newcommand{\FC}{{\boldsymbol{\mathcal{F}}}}
\newcommand{\HC}{{\boldsymbol{\mathcal{H}}}}
\newcommand{\LC}{{\boldsymbol{\mathcal{L}}}}
\newcommand{\MC}{{\boldsymbol{\mathcal{M}}}}
\newcommand{\NC}{{\boldsymbol{\mathcal{N}}}}
\newcommand{\OC}{{\boldsymbol{\mathcal{O}}}}
\newcommand{\QC}{{\boldsymbol{\mathcal{Q}}}}
\newcommand{\VC}{{\boldsymbol{\mathcal{V}}}}
\newcommand{\XC}{{\boldsymbol{\mathcal{X}}}}
\newcommand{\YC}{{\boldsymbol{\mathcal{Y}}}}
\newcommand{\ZC}{{\boldsymbol{\mathcal{Z}}}}
\newcommand{\DCo}{\DC_{\!\circ}}

\newcommand{\DCh}{{\rlap{\;/}\DC}}
\newcommand{\uDCh}{{\rlap{\;/}{\ul\DC}}}
\newcommand{\DCho}{{\rlap{\;/}\DCo}}

\newcommand{\VCo}{{\VC\!_{\circ}}}
\newcommand{\uVC}{\underline\VC}
\newcommand{\uVClin}{\underline\VC^*{}}
\newcommand{\uZC}{\underline\ZC}
\newcommand{\uZClin}{\underline\ZC^*{}}
\newcommand{\End}{\operatorname{End}}
\newcommand{\Aut}{\operatorname{Aut}}

\newcommand{\Tr}{\operatorname{Tr}}
\newcommand{\Ker}{\operatorname{Ker}}
\newcommand{\ad}{\operatorname{ad}}

\newcommand{\Id}[1]{{1\!\!1}\!{}_{#1}{}}
\newcommand{\id}{{1\!\!1}}
\newcommand{\dO}{\mathrm{d}}
\newcommand{\dH}{\mathrm{d}_{\sst{\mathrm H}}}
\newcommand{\DO}{\mathrm{D}}
\newcommand{\HO}{\mathrm{H}}
\newcommand{\jO}{\mathrm{j}}
\newcommand{\JO}{\mathrm{J}}
\newcommand{\TO}{\mathrm{T}}
\newcommand{\TS}{\TO^{*}\!}
\newcommand{\VO}{\mathrm{V}}
\newcommand{\VS}{\VO^{*}\!}
\newcommand{\eO}{\mathrm{e}}
\newcommand{\iO}{\mathrm{i}}
\newcommand{\pr}[1]{\operatorname{pr}_{#1}}
\newcommand{\na}{\nabla\!}
\newcommand{\nasl}{{\rlap{\raise1pt\hbox{\,/}}\nabla}}
\newcommand{\ten}[1]{\operatorname*{\otimes}_{\!{\scriptscriptstyle #1}} }
\newcommand{\cart}[1]{\operatorname*{\times}_{\!{\scriptscriptstyle #1}} }
\newcommand{\dir}[1]{\operatorname*{\oplus}_{\!{\scriptscriptstyle #1}} }
\newcommand{\we}{{\,\wedge\,}}
\newcommand{\weu}[1]{{\wedge^{\!#1}}}
\newcommand{\mdots}{{\cdot}{\cdot}{\cdot}}
\newcommand{\pint}{\mathord{\rfloor}}
\newcommand{\comp}{\mathbin{\raisebox{1pt}{$\scriptstyle\circ$}}}
\newcommand{\tn}{{\,\otimes\,}}

\newcommand{\bang}[1]{{\langle#1\rangle}}
\newcommand{\bra}[1]{{\langle#1|}}
\newcommand{\ket}[1]{{|#1\rangle}}
\newcommand{\Ii}[2]{{}^{\smash{#1}}_{\phantom{\smash{#1}}\!\smash{#2}}}
\newcommand{\iI}[2]{{}_{\smash{#1}}^{\phantom{\smash{#1}}\!\smash{#2}}}
\newcommand{\iIi}[3]{{}_{\smash{#1}\phantom{\smash{#2}}\!\!\smash{#3}}^{\phantom{\smash{#1}}\!\smash{#2}}}
\newcommand{\IiI}[3]{{}^{\smash{#1}\phantom{\smash{#2}}\!\!\smash{#3}}_{\phantom{\smash{#1}}\!\smash{#2}}}
\newcommand{\sA}{{\scriptscriptstyle A}}
\newcommand{\sB}{{\scriptscriptstyle B}}
\newcommand{\sH}{{\scriptscriptstyle H}}
\newcommand{\sI}{{\scriptscriptstyle I}}
\newcommand{\sJ}{{\scriptscriptstyle J}}
\newcommand{\Bsf}{{\mathsf{B}}}
\newcommand{\bb}{{\mathsf{b}}}
\newcommand{\Ksf}{{\mathsf{K}}}
\newcommand{\Xsf}{{\mathsf{X}}}
\newcommand{\ff}{{\mathsf{f}}}
\newcommand{\p}{{\scriptstyle\Pi}}
\newcommand{\uu}{{\mathsf{u}}}
\newcommand{\vv}{{\mathsf{v}}}
\newcommand{\sref}[1]{\S\ref{#1}}
\newcommand{\ie}{i.e$.$}
\newcommand{\eg}{e.g$.$}
\newcommand{\sst}{\scriptscriptstyle}

\newcommand{\into}{\hookrightarrow}
\newcommand{\onto}{\rightarrowtail}
\newcommand{\ul}{\underline}
\newcommand{\gh}{\omega}
\newcommand{\agh}{\varpi}
\newcommand{\ghost}{{}_{\sst\mathrm{ghost}}}
\newcommand{\spec}[1]{{}_{\sst{\mathrm{#1}}}}
\newcommand{\abs}[1]{{\mathrm{#1}}}
\newcommand{\cre}[1]{\abs{#1}^{\dag}}
\newcommand{\tra}[1]{\abs{#1}^{\!*}}

\newcommand{\brstS}{{\scriptstyle{\mathrm S}}}
\newcommand{\brstQ}{{\scriptstyle{\mathrm Q}}}
\newcommand{\suc}[1]{{\{\![#1]\!\}}}
\newcommand{\Suc}[1]{{\bigl\{\!\!\bigl[#1\bigr]\!\!\bigl\}}}
\newcommand{\swe}{\,{\scriptstyle\lozenge}\,}
\newcommand{\grade}[1]{{\lfloor#1\rceil}}

\newcommand{\scc}{{\mathsf{c}}}
\newcommand{\sco}[3]{\scc\Ii{#1}{#2#3}}
\newcommand{\bwe}{\,{\barwedge}\,}

\newcommand{\ssGa}{{\scriptscriptstyle\Gamma}}
\title{Fr\"olicher-smooth geometries,
quantum jet bundles \\ and BRST symmetry}
\date{{\small December 1, 2014} }
\author{Daniel Canarutto\\[6pt]
{\small\it Dipartimento di Matematica e Informatica ``U.~Dini'', }\\
{\small\it Via S. Marta 3, 50139 Firenze, Italia}\\
{\small email:~daniel.canarutto@unifi.it}\\
{\small http://www.dma.unifi.it/\char126 canarutto}}
\begin{document}
\bibliographystyle{alpha}
\maketitle
\begin{abstract}\noindent
We attempt a clarification of geometric aspects of quantum field theory
by using the notion of smoothness introduced by Fr\"olicher
and exploited by several authors in the study of functional bundles.
A discussion of momentum and position representations in curved spacetime,
in terms of generalized semi-densities,
leads to a definition of quantum configuration bundle
which is suitable for a treatment of that kind.
A consistent approach to Lagrangian field theories,
vertical infinitesimal symmetries and related currents is then developed,
and applied to a formulation of BRST symmetry in a gauge theory
of the Yang-Mills type.
\end{abstract}

\bigbreak

\bigbreak

\bigbreak
\noindent
{\sc Journal reference}:\\
D.\ Canarutto:
`Fr\"olicher-smooth geometries, quantum jet bundles and BRST symmetry',\\
J.\ Geom.\ Phys.\ (2014), http://dx.doi.org/10.1016/j.geomphys.2014.11.013.

\vfill\newpage
\tableofcontents
\vfill\newpage

\thispagestyle{empty}
\section*{Introduction and summary}

Our understanding of quantum field theory (QFT) may benefit from
clarifications regarding the differential geometric notions underlying it.
Usually one considers a quantised version of a classical field,
which is a section of a finite-dimensional bundle over spacetime,
by seeing it as ``operator-valued''.
Apparently, the standard formalism of differential geometry
seamlessly carries over to this infinite-dimensional setting,
but a closer look reveals complications.

One should also bear in mind that QFT
is presently better understood and effective in a very specialized context,
where one essentially works in Minkowski spacetime with a chosen observer,
and background parallel transport of internal configurations is available.
These background structures allow various simplifications and identifications;
in particular, the spatial Fourier transform yields a full correspondence
between position and momentum representations.
A possible covariant formulation on a curved fixed background,
on the other hand,
besides giving us some opportunity
of studying gravitational effects in particle physics,
constrains our setting to be meaningful at a higher level,
and also allows us to distinguish where exactly
the various needed assumptions enter the picture.

A useful ingredient in that quest is the notion
of smoothness introduced by Fr\"olicher~\cite{Fr},
or \emph{F-smoothness}, which has been studied and exploited by several
authors~\cite{FK,KrieglMichor97,CK95,CK97,CK99,CJK04,CJK06,KolarModugno98,C04a}
and provides a convenient approach to infinite-dimensional bundles.
The differential geometric notions
of tangent, vertical and jet functors, bundle connections,
the Fr\"olicher-Nijenhuis algebra of tangent-valued forms,
all can be introduced in the context of functional and distributional bundles
over a finite dimensional base manifold
by a fairly direct and general procedure,
without heavy involvement in infinite-dimensional topology.
In particular, the notion of a smooth connection of a functional bundle
has been applied in the context of the ``covariant quantization''
approach to Quantum Mechanics~\cite{CJM,JM02,JM00_02,JM06,TePrVi14,Vi99,Vi00},
while the geometry of distributional bundles is used in an approach,
proposed by this author,
to the physics of interacting quantum particles~\cite{C05,C12a}.

In the literature one finds works about BRST cohomology and related issues
which treat the matter in a much more ample and general way
than this one~\cite{Brandt97,BBH00,Brandt01,BGMS05,BGMS07}.
Admittedly we do not aim at such generality,
but rather at a close examination of some essential notions
in the concrete context of Yang-Mills theories.
Thus we try to avoid intricacies such as infinite jets
or Fr\'echet space topology,
while at the same time we try not to overlook certain complications
which, particularly in the physical literature,
are often dealt with in a rather informal way.

F-smoothness enters the present paper in two distinct ways:
the geometry of quantum state bundles,
which was introduced in previous papers and is reviewed here,
and the geometry of those bundles over spacetime
whose sections are the quantum fields.
The relation between these two complementary aspects is examined
in~\sref{s:Quantum bundles and quantum fields},
and it is argued that their equivalence is, in general,
only partial and dependent on a suitable linearization
and the choice of a spacetime synchronization;
but we remark that a distinguished time
is actually needed in all approaches to QFT.

In~\sref{s:F-smooth geometry and Lagrangian field theory}
we lay down the basics of F-smooth geometry for quantum fields.
The quantum configuration bundle is obtained from the classical bundle
via fiber tensor product by a certain infinite dimensional $\ZZ_2$-graded algebra which was defined in~\sref{s:Quantum bundles and quantum fields}.
Jet bundles and connections can be introduced by a straightforward adaptation
of the general procedure already used in other contexts,
while the notion of fiber partial derivative
and dual bundles require special care.
We then apply those ideas to a jet-bundle approach to Lagrangian field theory,
drawing the guidelines from the well-developed classical theory~%
\cite{HorakKolar83,Kolar84,MaMo83b,KrupkaKrupkovaSanders10,KrupkaSanders08,
VinogradovCSS84I,VinogradovCSS84II}.
There one deals with arbitrary exterior forms
on jet bundles of the configuration bundle,
which can be actually introduced in the said quantum context
but present us with some complications.
In order to deal with \emph{vertical} infinitesimal symmetries, however,
the restricted notion of \emph{basic form}
(or ``totally horizontal'' form \cite{Sa89})
is sufficient.
The symmetry naturally acts on such forms by a ``pseudo-Lie'' derivation
which raises the order of the form.
The result of applying this to a Lagrangian density
splits as a horizontal differential plus (essentially)
the Euler-Lagrange operator,
whence we derive a special version of the Noether theorem
and the notions of current and charge associated with the symmetry.

Finally we apply these ideas to a formulation of BRST symmetry
in a gauge theory of the Yang-Mills type.
The classical geometric background and the issue of gauge freedom
are discussed in a fairly general setting,
though we provisionally disregard the distinction
between the right-handed and left-handed sectors and symmetry breaking.
Then we introduce the vertical vector field generating the symmetry,
and calculate the corresponding current.
A comparison of the equivalent first-order and second-order versions
of the ghost Lagrangian yields some insight
about Noether's theorem in this context.

\section{Quantum bundles and quantum fields}
\label{s:Quantum bundles and quantum fields}
\subsection{Finitely-generated multi-particle algebra}
\label{ss:Finitely-generated multi-particle algebra}

Let \hbox{$\Z\onto\X$} be a finite-dimensional vector bundle.
Denote by $\uZC^1$ the vector space of all sections \hbox{$z:\X\to\Z$}
which vanish outside some finite subset \hbox{$\X\!_z\subset\X$},
$$\uZC^1\equiv
\Bigl\{\ket{z}=\sum_{x\in\X\!_z} \ket{z(x)}\Bigr\}~,\quad
\X\supset\X\!_z~\text{finite}$$
(we denote $z$ and \hbox{$z(x)\in\Z_x$} as ``ket'' objects
when it is convenient).
Note that the term ``section'' is used here in a broad sense,
and $\X$ could actually be an arbitrary set;
in practice, we'll deal with smooth bundles.

This $\uZC^1$ is our template for the space of states
of one particle of some type.
We define the associated ``$n$-particle state'' space $\uZC^{n}$
to be either the symmetrised tensor product $\vee^n\uZC^1$ (\emph{bosons})
or the  anti-symmetrised tensor product $\weu{n}\uZC^1$ (\emph{fermions}).
The ``multi-particle state'' space is
\hbox{$\uZC\equiv\bigoplus_{n=0}^\infty\uZC^{n}$}
(constituted by finite sums with arbitrarily many terms).
If \hbox{$y\in\uZC^{m},\,z\in\uZC^{n}$},
then we define \hbox{$y\swe z\in\uZC^{m+n}$} to be either
$y\,{\vee}\,z$ or $y\we z$\,.
By abuse of language we call this the ``exterior product'' of $y$ and $z$\,,
and extend it to any elements in $\uZC$ by linearity.

Let \hbox{$\Z^*\onto\X$} be the dual vector bundle.
The dual space of $\uZC^1$ is the vector space
of \emph{all} sections \hbox{$\X\to\Z^*$},
which we can formally write as infinite sums
\hbox{$\z=\sum_{x\in\X} \bra{\z(x)}$}.
Actually
\hbox{$\bang{\z,z}=\sum_{x\in\X\!_z}\bang{\z(x),z(x)}$}, a finite sum.
For our purposes we may as well work with its subspace \hbox{$\uZClin^1$}
constituted of all such sections which vanish outside some finite subset.
Exterior product (with the same parity as that of the associated $\uZC^1$)
yields now spaces \hbox{$\uZClin^{n}$} and the ``dual multi-particle space''
\hbox{$\uZClin\equiv\bigoplus_{n=0}^\infty\uZClin^{n}$}.

Next we introduce an ``interior product'' \hbox{$\l\,|\,\psi$}\,,
where \hbox{$\psi\in\uZC{}$} and \hbox{$\l\in\uZClin{}$}.
For fermions, this is the usual interior product
\hbox{$i[\l]\psi$} of exterior algebra.
For bosons it can be defined similarly,
as tensor contraction with appropriate symmetrization and normalization,
so that the rule \hbox{$(\z\swe\l)\,|\,\psi=\l\,|\,(\z\,|\,\psi)$} holds
for all \hbox{$\z\in\uZClin^1$}, \hbox{$\l\in\uZClin$}.

A general theory of quantum particles has several particle types.
Correspondingly, one considers several multi-particle state spaces
(or ``sectors'')
$\uZC'$, $\uZC''$, $\uZC'''$ etc.
The total state space is now defined to be
$$\uVC:=\uZC'\tn\uZC''\tn\uZC'''\tn\mdots=
\textstyle{\bigoplus_{n=0}^\infty}\uVC^{n}$$
where $\uVC^{n}$, constituted of all elements of tensor rank $n$\,,
is the space of all states of $n$ particles of any type.
We observe that if $\XC$ and $\YC$ are any two vector spaces,
then their antisymmetric tensor algebras fulfill the isomorphisms
$$\weu{p}(\XC\,{\oplus}\,\YC)\cong
\textstyle{\bigoplus_{h=0}^p}\,(\weu{p-h}\XC)\tn(\weu{h}\YC)~,\quad
(\wedge\XC)\tn(\wedge\YC)\cong\wedge(\XC\,{\oplus}\,\YC)~.$$
Hence all fermionic sectors can be described by a unique
overall antisymmetrised tensor algebra.
A similar observation holds true for the bosonic sectors,
while we regard mutual ordering of fermionic and bosonic sectors as inessential.
Similarly one constructs a ``dual'' space
\hbox{$\uVClin:=\uZClin'\tn\uZClin''\tn\uZClin'''\mdots=
\bigoplus_{n=0}^\infty\uVClin^{n}$}.

If we now let the \emph{grade} $\grade{\phi}$
of a monomial element (a ``decomposable tensor'') \hbox{$\phi\in\VC$}
to be the parity of the number of fermion factors it contains,
then we can see $\uVC$ as a ``super-algebra'' (a $\ZZ_2$-graded algebra),
products being performed in the appropriate tensor factors.
Similarly, the interior product is defined by performing
interior products in the appropriate tensor factors.
We then obtain the rules
\begin{align*}
&\psi\swe\phi=(-1)^{\grade{\phi}\grade{\psi}}\phi\swe\psi~,\quad
(\z\swe\xi)\,|\,\psi=\xi\,|\,(\z\,|\,\psi)~,
\\[6pt]
&z\,|\,(\phi\swe\psi)=
(z\,|\,\phi)\swe\psi+(-1)^{\grade{z}\grade{\phi}}\,\phi\swe(z\,|\,\psi)~,\qquad
\phi,\psi\in\VC,~
\z,\xi\in\uVClin^1,
\end{align*}
valid whenever each of the involved factors has a definite grade.
A linear map \hbox{$X:\uVC\to\uVC$} is called
a \emph{super-derivation} (or \emph{anti-derivation}) \emph{of grade~$\grade{X}$}
if \hbox{$\grade{X\psi}=\grade{X}+\grade{\psi}$}
and if it fulfills the graded Leibnitz rule\footnote{
This expression, as well as others which follow,
involves decomposable elements;
extension by linearity must be understood.}
$$X(\phi\swe\psi)=(X\phi)\swe\psi+
(-1)^{\grade{X}\grade{\phi}}\phi\swe X\psi~.$$

The \emph{absorption} operator associated with \hbox{$\z\in\uVClin^1$}
and the \emph{emission} operator associated with \hbox{$z\in\uVC^1$}
are the linear maps \hbox{$\uVC\to\uVC$} respectively defined as
$$\abs{a}[\z]\phi\equiv \z\,|\,\phi~,\quad
\tra{a}[z]\phi\equiv z\swe\phi~,\qquad\phi\in\uVC~.$$
Similarly we have operators
\hbox{$\abs{a}[z],\tra{a}[\z]:\uVClin\to\uVClin$}.
We obtain \hbox{$\l\,|\,\abs{a}[\z]\psi=(\tra{a}[\z]\l)\,|\,\psi$}
for all \hbox{$\l\in\uVClin$},
namely $\abs{a}[\z]$ and $\tra{a}[\z]$
are mutually transposed endomorphisms.

By finite compositions and finite linear combinations,
absorption and emission operators generate a vector subspace
\hbox{$\ul\OC\subset\End\uVC$}.
This turns out to be a $\ZZ_2$-graded algebra
(the algebra product being the composition of endomorphisms)
by letting the grades of $\abs{a}[\z]$ and $\tra{a}[z]$
be $\grade{\z}$ and $\grade{z}$\,, respectively.
The \emph{super-bracket} of \hbox{$X,Y\in\ul\OC$} is then defined by
$$\suc{X,Y}:=X\,Y-(-1)^{\sst\grade{X}\grade{Y}}Y\,X$$
whenever both $X$ and $Y$ have definite grade, and extended by linearity.
In particular, for all \hbox{$y,z\in\uVC^1$}
and \hbox{$\z,\xi\in\uVClin^1$} we get
$$\suc{\abs{a}[\xi],\abs{a}[\z]}=\suc{\tra{a}[y],\tra{a}[z]}=0~,\quad
\suc{\abs{a}[\z],\tra{a}[z]}=\bang{\z,z}\,\id~.$$

\remark~If $X$ and $Y$ are derivations then $\suc{X,Y}$ turns out to be a derivation
of grade \hbox{$\grade{X}+\grade{Y}$}.
Furthermore, the map
\hbox{$\ad_{\sst X}:Y\mapsto\ad_{\sst X}\!Y\equiv\suc{X,Y}$}
turns out to be a derivation of grade
\hbox{$\grade{\ad_{\sst X}}=\grade{X}$} in $\ul\OC$\,.
\smallbreak

On the other hand we also have, on the same underlying vector space,
a graded algebra
\hbox{$\ul\OC\equiv\bigoplus_{n=0}^\infty\ul\OC^{n}$}
where \hbox{$\ul\OC^0\equiv\CC$} and $\ul\OC^n$
is the space spanned by compositions of $n$ emission and absorption operators.
Now we observe that \emph{as a vector space} $\ul\OC$ is naturally isomorphic to
\hbox{$\uVC\tn\uVClin\cong\lozenge_{n=0}^\infty(\uVC^1\oplus\uVClin^1)$}\,.
The identification is obtained by using the above super-commutation rules
in order to move all absorption operators to the right of any emission operators
(\emph{normal order}).
On the other hand, if we replace the last rule by letting
$\abs{a}[\z]$ and $\tra{a}[z]$ super-commute,
then we get an isomorphism of $\NN$-graded algebras.
The situation is quite similar to that of a Clifford algebra,
where the scalar product determines a new algebra structure
on the exterior algebra of a vector space;
the new product is not ``super-commutative''.

In most practical cases \hbox{$\Z\onto\X$} is a complex bundle
with a Hermitian structure in its fibers.
This yields an isomorphism \hbox{$\#:\Z^*\to\Zc:\z\mapsto\z^\#$} over $\X$,
where \hbox{$\Zc\onto\X$} is the conjugate bundle;
the inverse of $\#$ is denoted as \hbox{$\flat:\bz\mapsto\bz^\flat$}.
The induced anti-isomorphism \hbox{$\uZC^1\to\uZClin^1$} is traditionally
denoted as \hbox{$\ket{z}\mapsto\bra{z}$}\,.
Extending it to an anti-isomorphism
\hbox{$\uZC\leftrightarrow\uZClin$} is straightforward.

We also note that whenever a sector corresponding
to a complex bundle $\Z$ is considered,
then the theory also includes the sector corresponding to $\Zc$.
Accordingly, for \hbox{$\z\in\uVClin^1$}
we have the ``anti-particle'' emission operator
\hbox{$\cre{a}[\z]\equiv\tra{a}[\z^\#]$}\,.

\begin{remark}
In the case of Dirac spinors
one uses two different Hermitian structures:
the (intrinsic) Dirac conjugation is denoted as \hbox{$z\mapsto\bz^\flat$}
(usually just $\bz$), while
the conjugation determined by the \emph{positive} Hermitian structure
associated with the chosen observer
is denoted as \hbox{$z\mapsto z^\dag\equiv\bz^\flat\comp\g^0$}\,.
\end{remark}

\subsection{Quantum states as generalised semi-densities}
\label{ss:Quantum states as generalised semi-densities}

Let \hbox{$\Z\onto\X$} be a finite-dimensional complex vector bundle,
\hbox{$\dim_\RRr\X=m$}\,.
Assume that $\X$ is \emph{orientable}, and choose a positive
semi-vector bundle\footnote{
For an account of positive semi-spaces and their rational powers,
see~\cite{C12a} and the bibliography therein.}
\hbox{$\VV\equiv\VV\X\equiv(\weu{m}\TO\X)^+$}\,.
A section \hbox{$\X\to\VV^{-1/2}\tn\Z$} is called
a \emph{$\Z$-valued semi-density}.
The vector space of all such sections which are smooth and have compact support
is denoted as $\DCho(\X,\Z)$\,.
The dual space of $\DCho(\X,\Z^*)$ in the standard test map topology~\cite{Sc}
is indicated as \hbox{$\DCh(\X,\Z)$} and called the space of
\emph{$Z$-valued generalised semi-densities}.
In particular, a sufficiently regular ordinary section
\hbox{$\th:\X\to\VV^{-1/2}\tn\Z$} is in $\DCh(\X,\Z)$ via the rule
\hbox{$\bang{\th,\s}:=\int_\X\!\bang{\th(x),\s(x)}$}\,,
\hbox{$\s\in\DCho(\X,\Z^*)$}\,.

Semi-densities have a special status among all kinds of generalised sections
because of the natural inclusion $\DCho(\X,\Z)\subset\DCh(\X,\Z)$\,.
Furthermore, if a fibered Hermitian structure
of \hbox{$\Z\onto\X$} is assigned then one has the space $\LC^2(\X,\Z)$
of all ordinary semi-densities $\th$ such that
\hbox{$\bang{\th^\dag,\th}<\infty$}\,.
The quotient \hbox{$\HC(\X,\Z)=\LC^2(\X,\Z)/\boldsymbol{0}$}
is then a Hilbert space
(here \hbox{$\boldsymbol{0}\subset\LC^2(\X,\Z)$} denotes the subspace
of all almost-everywhere vanishing sections),
and we get a so-called \emph{rigged Hilbert space}~\cite{BLT}
$$\DCho(\X,\Z)\subset\HC(\X,\Z)\subset\DCh(\X,\Z)~.$$
Elements in \hbox{$\DCh(\X,\Z)\setminus\HC(\X,\Z)$} can then be identified
with the (\emph{non-normalizable}) \emph{generalised states}
of the common terminology.

Let $\d[x]$ be the \emph{Dirac density} on $\X$
with support $\{x\}$\,, \hbox{$x\in\X$}\,.
A generalised semi-density is said to be \emph{of Dirac type}
if it is of the form \hbox{$\d[x]\tn v\in\DCh(\X,\Z)$}
with \hbox{$v:\X\to\VV^{1/2}\tn\Z$}.
We define $\uDCh(\X,\Z)$ to be the space of all
\emph{finite linear combinations} of Dirac-type semi-densities.
An important result in the theory of distributions~\cite{Sc}
implies that $\uDCh(\X,\Z)$ is dense in $\DCh(\X,\Z)$,
namely any generalised semi-density can be approximated
with arbitrary precision
(in the sense of the topology of distributional spaces)
by a finite linear combination of Dirac-type semi-densities.

An isomorphism
\hbox{$\uZC^1\leftrightarrow\uDCh^1\equiv\uDCh(\X,\Z)$}
is determined by the assignment of a volume form \hbox{$\eta:\X\to\VV^{-1}$}.
If $\bigl(\bb_\a\bigr)$ is a frame of \hbox{$\Z\onto\X$}
(for notational simplicity we assume its domain to be the whole $\X$),
then this isomorphism is characterised by the correspondence
$$\ket{x}\tn\bb_\a(x) \leftrightarrow
\Bsf_{x\a}\equiv\d[x]\tn\eta^{-1/2}\tn\bb_\a(x)~.$$

The set \hbox{$\bigl(\Bsf_{x\a}\bigr)\subset\DCh^1$}
is called a \emph{generalised basis}.
Accordingly we introduce a handy ``generalised index'' notation.
We write \hbox{$\Bsf^{x\a}\equiv\d[x]\tn\eta^{-1/2}\tn\bb^\a(x)$}\,,
where $\bigl(\bb^\a\bigr)$ is the dual classical frame.
Though contraction of any two distributions is not defined in the ordinary sense,
a straightforward extension of the discrete-space operation yields
$$\bang{\Bsf^{x'\!\a'},\Bsf_{x\a}}=\d^{x'}_x\,\d^{\a'}_{\a}~,$$
where $\d^{x'}_x$ is the generalised function
usually indicated as $\d(x'\,{-}\,x)$\,.
This is consistent with ``index summation'' in a generalised sense:
if \hbox{$z\in\DCho(\X,\Z)$}
and \hbox{$\z\in\DCho(\X,\Z^*)$} are test semi-densities,
then we write
\begin{align*}
&z^{x\a}\equiv z^\a(x)\equiv\bang{\Bsf^{x\a},z}~,\quad
\z_{x\a}\equiv \z_\a(x)\equiv\bang{\z,\Bsf_{x\a}}~,
\\[6pt]
&\bang{\z,z}\equiv
\z_{x'\!\a'}\,z^{x\a}\,\bang{\Bsf^{x'\!\a'},\Bsf_{x\a}}\equiv
\int_\X \z_\a(x)\,z^\a(x)\,\eta(x)~,
\end{align*}
namely we interpret index summation
with respect to the continuous variable $x$ as integration,
provided by the chosen volume form.
This formalism can be extended to the contraction
of two generalised semi-densities whenever it makes sense.

We extend the above constructions to multi-particle spaces.
The identification \hbox{$\uZC^1\leftrightarrow\uDCh^1$}
extends to \hbox{$\uZC^n\leftrightarrow\uDCh^n\equiv\lozenge^n\uDCh^1$}\,.
Then $\uZC^n$ turns out to be dense in $\ZC^n$,
defined to be either the symmetrised or the antisymmetrised subspace of
$\DCh(\X^n,{\otimes}^n\Z)$\,,
\hbox{$\X^n\equiv\X\,{\times}\,\mdots\,{\times}\,\X$}\,.
Next we set \hbox{$\ZC\equiv\bigoplus_{n=0}^\infty\ZC^{n}$},
and assemble several particle types into one total state space
\hbox{$\VC:=\ZC'\tn\ZC''\tn\mdots$}.
An analogous construction yields the ``dual'' space $\VC^*$.
Furthermore, using the spaces $\DCho$ of test semi-densities
we obtain subspaces \hbox{$\VCo\subset\VC$}
and \hbox{$\VC\!_{\circ}^{\:{*}}\subset\VC^*$}.
All these spaces are naturally $\ZZ_2$-graded.

Also the constructions related to the operator algebra
(\sref{ss:Finitely-generated multi-particle algebra})
can be extended to the present setting.
If \hbox{$\z\in\VC^{*1}$} and \hbox{$z\in\VC^1$}
then the absorption operator $\abs{a}[\z]$
and the emission operator $\tra{a}[z]$
are well-defined linear maps \hbox{$\VCo\to\VC$}\,.
The vector space $\OC^1$ of all sums of the kind \hbox{$\abs{a}[\z]+\tra{a}[z]$}
has the subspace $\ul\OC^1$ of all finite linear combinations of
absorption and emission operators associated with Dirac-type semi-densities.
In particular we write \hbox{$\abs{a}^{x\a}\equiv\abs{a}[\Bsf^{x\a}]$}\,,
\hbox{$\tra{a}_{x\a}\equiv\tra{a}[\Bsf_{x\a}]$}\,,
and obtain super-commutation rules
$$\Suc{\abs{a}^{x\a},\abs{a}^{x'\!\a'}}=
\Suc{\tra{a}_{x\a},\tra{a}_{x'\!\a'}}=0~,\quad
\Suc{\abs{a}^{x\a},\tra{a}_{x'\!\a'}}=\d^\a_{\a'}\,\d^x_{x'}~,$$
where the latter is to be understood in a generalised sense:
for \hbox{$\z\in\VC\!_{\circ}^{\;{*}1}$}, \hbox{$z\in\VC\!_{\circ}^{\;1}$}\,,
we write
$$\Suc{\abs{a}[\z],\tra{a}[z]}=
\Suc{\z_{x\a}\,\abs{a}^{x\a},z^{x'\!\a'}\,\tra{a}_{x'\!\a'}}=
\z_{x\a}\,z^{x'\!\a'}\,\Suc{\abs{a}^{x\a},\tra{a}_{x'\!\a'}}=
\bang{\z,z}~.$$

Next we denote as $\OC^n$, \hbox{$n\in\NN$}\,,
the vector space spanned by all compositions
of normally ordered $n$ emission and absorption operators.
A product \hbox{$\OC^n\times\OC^p\to\OC^{n+p}$} can be defined as composition
together with normal reordering,
obtained by imposing the \emph{modified rule}
\begin{equation}\label{equation:modifiedsupercommutator}
\Suc{\abs{a}^{x\a},\tra{a}_{x'\!\a'}}=0~.
\end{equation}
Setting \hbox{$\OC^0\equiv\CC$}
we obtain a graded algebra \hbox{$\OC\equiv\bigoplus_{n=0}^\infty\OC^{n}$}
of linear maps \hbox{$\VCo\to\VC$}
(we remark that here,
differently from~\sref{ss:Finitely-generated multi-particle algebra},
\emph{normal order is needed} for obtaining an algebra of such maps).
Moreover, $\OC$ turns out to be a $\ZZ_2$-graded algebra
by letting the grades of $\abs{a}[\z]$ and $\tra{a}[z]$
be $\grade{\z}$ and $\grade{z}$\,, and we have the isomorphism
\hbox{$\OC\cong\VC\tn\VC^*$}.

Let \hbox{$Z:\RR\to\OC$} be a local curve such that
\hbox{$\lim_{\l\to0}[Z(\l)\chi]\in\VC$} exists in the sense of distributions
for all \hbox{$\chi\in\VCo$}\,.
Then $\lim_{\l\to0}Z(\l)$ is a well-defined linear map \hbox{$\VCo\to\VC$}
which belongs, in general, to an extended space \hbox{$\OC^\bullet\supset\OC$}\,.
In this paper we won't be concerned with the issue
of defining a topology on $\OC^\bullet$,
as the notion of F-smoothness will suffice for our purposes.

\subsection{Fr\"olicher-smoothness}
\label{ss:Froelicher-smoothness}

Let $\MC$ be any set.
A family $\CCal\!_{\sst\MC}$ of curves $\RR\to\MC$
determines a family $\FC\CCal\!_{\sst\MC}$ of maps $\MC\to\RR$ by the rule
$$f\in\FC\CCal\!_{\sst\MC}\iff
f\comp c\in\CCal^\infty(\RR)~\forall\,c\in\CCal\!_{\sst\MC}~.$$
Conversely, a set $\FC\!\!_{\sst\MC}$ of functions $\MC\to\RR$ determines
a set $\CCal\FC\!\!_{\sst\MC}$ of curves in $\MC$ by
$$c\in\CCal\FC\!\!_{\sst\MC}\iff
f\comp c\in\CCal^\infty(\RR)~\forall\,f\in\FC\!\!_{\sst\MC}~.$$
An \emph{F-smooth structure} on $\MC$
is defined to be a couple $(\CCal\!_{\sst\MC},\FC\!\!_{\sst\MC})$ such that
$\CCal\!_{\sst\MC}$ and $\FC\!\!_{\sst\MC}$ determine each other,
namely \hbox{$\FC\CCal\!_{\sst\MC}=\FC\!\!_{\sst\MC}$}
and \hbox{$\CCal\FC\!\!_{\sst\MC}=\CCal\!_{\sst\MC}$}\,.
Note that either any set $\CCal_0$ of curves in $\MC$,
or any set $\FC_0$ of functions on $\MC$,
\emph{generate} an F-smooth structure either by
\hbox{$\FC\!\!_{\sst\MC}:=\FC\CCal_0$} or by
\hbox{$\CCal\!_{\sst\MC}:=\CCal\FC_0$}\,.
If \hbox{$(\NC,\CCal\!_{\sst\NC}\,,\FC\!\!_{\sst\NC})$}
is another F-smooth structure,
then a map \hbox{$\Phi:\MC\to\NC$} is called F-smooth if
\hbox{$\Phi\comp c\in\CCal_{\sst\NC}$} for all \hbox{$c\in\CCal\!_{\sst\MC}$}\,,
or equivalently if \hbox{$f\comp\Phi\in\FC\!\!_{\sst\MC}$}
for all \hbox{$f\in\FC\!\!_{\sst\NC}$}\,.

It can be proved~\cite{Bo} that a function
\hbox{$f:\M\to\RR$} on a classical manifold $\M$
is smooth (in the standard sense) if and only if the composition
\hbox{$f\comp c$} is a smooth function of one variable
for any smooth curve \hbox{$c:\RR\to\M$}\,.
Thus one has a unified notion of smoothness based on smooth curves,
including classical manifolds as well as functional and distributional spaces.
This notion of smoothness,
which was introduced by Fr\"olicher~\cite{Fr},
behaves naturally with regard to inclusions and cartesian products,
so it yields a convenient general setting for dealing
with functional spaces and functional bundles~%
\cite{FK,KrieglMichor97,CK95,KolarModugno98,C04a}.

\subsection{Fr\"olicher-smooth quantum state bundles}
\label{ss:Froelicher-smooth quantum state bundles}

F-smoothness in a distributional space is defined quite naturally:
a curve $c$ is called F-smooth if \hbox{$t\to\bang{c(t),u}$}
is smooth for any test element $u$\,.
Accordingly, \emph{distributional bundles}
are defined to be F-smooth vector bundles, over a classical base manifold,
whose fibers are distributional spaces.
The geometry of these bundles, and in particular the connections on them,
have been studied in a previous paper~\cite{C04a}.

Bundles of generalised semi-densities over particle momenta
(``quantum bundles'') are a special case,
which has been applied to a partly original approach
to quantum particle physics~\cite{C05}.
For a given particle type,
the underlying ``classical'' (\ie\ finite-dimensional)
geometric structure is that of a 2-fibered bundle
\hbox{$\Z\to\Pm\to\M$}
(the top fibers describing the `internal degrees of freedom'
of the considered particle type),
where $(\M,g)$ is Einstein's spacetime and \hbox{$\Pm\subset\P\cong\TS\M$}
is the sub-bundle over $\M$ of future shells for the particle's mass $m$
(4-momentum bundle).
At each \hbox{$x\in\M$} we perform the construction presented
in~\sref{ss:Quantum states as generalised semi-densities},
with the generic manifold $\X$ now replaced by $(\Pm)_x$\,.
We get spaces $\uZC^1_x$ and \hbox{$\ZC^1_x\equiv\DCh((\Pm)_x,\Z\!_x)$}\,.
The fibered sets \hbox{$\uZC^1:=\bigsqcup_{x\in\M}\!\!\uZC^1_x$}
and  \hbox{$\ZC^1:=\bigsqcup_{x\in\M}\!\!\ZC^1_x$}
turn out to have a natural F-smooth vector-bundle structure over $\M$.
Now the \emph{multi-particle state} bundles
$$\uZC:=\textstyle{\bigoplus_{n=0}^\infty}\uZC^{n}\onto\M~,\quad
\ZC:=\textstyle{\bigoplus_{n=0}^\infty}\ZC^{n}\onto\M~,$$
turn out to be  F-smooth vector bundles.
The same is true for the sub-bundles of $\ZC$
constructed from $\uDCh(\Pm,\Z)$\,, whose fibers are
constituted of finite linear combinations of Dirac-type semi-densities.
A natural isomorphism \hbox{$\uZC^1\leftrightarrow\uDCh(\Pm,\Z)$}
is determined by the mass-shell \emph{Leray form}.\footnote{
This is a distinguished volume form on the shells,
usually denoted as \hbox{$\d(p^2-m^2)$}.}
Considering more particle types, one eventually gets the total quantum bundle
$$\VC:=\ZC'\tn\ZC''\tn\ZC'''\tn\mdots=
\textstyle{\bigoplus_{n=0}^\infty}\VC^{n}\onto\M~.$$
Note that the quantum bundles for particle types of different mass
are constructed over different mass-shell bundles.

In order to proceed we must now consider an orthogonal splitting
\hbox{$\TS\M\equiv\P=\P_{\!\!\spar}\oplus\P_{\!\!\sbot}$}
into ``timelike'' and ``spacelike'' $g$-orthogonal subbundles over $\M$,
which can be seen as associated to the choice of an observer and is needed
in order to build a theory of quantum particles and their interactions.
Indeed, while we may wish we had an observer-independent theory,
we must accept that some kind of an observer and its proper time
are always used in particle physics at the present state of the art.
In~\sref{ss:Detectors and quantum configuration space}
we'll discuss the viability of a somewhat weaker requirement.

Let $\eta_\sbot$ be the volume form, associated with the metric,
on the fibers of \hbox{$\P_{\!\!\sbot}\onto\M$}.
The orthogonal projection
\hbox{$\P\to\P_{\!\!\sbot}$} yields a distinguished diffeomorphism
\hbox{$\Pm\leftrightarrow\P_{\!\!\sbot}$} for each $m$.
The pull-back of $\eta_\sbot$ is then
a volume form on the fibers of $\Pm$\,,
which is denoted for simplicity by the same symbol.
The Leray form can be then written as
\hbox{$\om_m(p)=(2\,p_0)^{-1}\eta_\sbot(p)$}\,, \hbox{$p\in\Pm$}\,,
where \hbox{$p_0\equiv\Eo_m(p_\sbo)=(m^2+p_\sbo^2)^{1/2}$}.

It will be convenient to use the ``spatial part''
$p_\sbo$ of the 4-momentum $p$ as a label,
that is a generalised index for quantum states.
For each \hbox{$p\in\Pm$} let $\d_m[p]$
the Dirac density with support $\{p\}$ on the same fiber of $\Pm$\,,
and \hbox{$\d(y_\sbo\!{-}p_\sbo)$} the generalised function characterised by
\hbox{$\d_m[p](y)=\d(y_\sbo\!{-}p_\sbo)\,\dO^3y$}
in terms of linear coordinates
\hbox{$\bigl(y_\l\bigr)\equiv\bigl(y_0,y_1,y_2,y_3\bigr)
\equiv\bigl(y_0,y_\sbo\bigr)$}
in the fibers of $\P$.
Now consider the section \hbox{$\Pm\to\DCh(\Pm,\CC):p\mapsto\Xsf_p$}
defined as follows;
for each \hbox{$p\in\Pm$} we can regard $\Xsf_p$
as a generalised function of the variable $y_\sbo$\,,
with the expression
$$\Xsf_p(y):=l^{-3/2}\,\d(y_\sbo{-}p_\sbo)\,\sqrt{\dO^3y}~.$$
Here $l$ is a constant length needed
in order to get an unscaled (``conformally invariant'') semi-density
(compare with the usual ``box quantization'' argument).
Eventually, we get the distinguished isomorphism
\hbox{$\uZC^1\leftrightarrow\uDCh(\Pm,\Z)$}
which is determined by the correspondence
\hbox{$\ket{z}\leftrightarrow\Xsf_p\tn z$}\,, \hbox{$z\in\Z_p$}\,.

The dual frame of $\bigl\{\Bsf_{p\a}\bigr\}$ is $\bigl\{\Bsf^{p\a}\bigr\}$ where
$$\Bsf^{p\a}=\Xsf^p\,\tn\bb^\a~.$$
Here, $\bigl(\bb^\a\bigr)$ is the classical dual frame of $\bigl(\bb_\a\bigr)$\,;
$\Xsf^p$, the dual of $\Xsf_p$\,, is actually the same semi-density.
We obtain
\hbox{$\bang{\Bsf^{p\a},\Bsf_{q\b}}=l^{-3}\,\d(p_\sbo-q_\sbo)\,\d^a_\b$}\,,
an unscaled relation.

Let \hbox{$h:\Pm\to\Za\tn\Z^*$} be the tensor
describing the Hermitian structure of $\Z$.
Then \hbox{$(\bb^\a)^\#=h^{\a\.\,\a}\,\bar\bb_{\a\.\,}:\Pm\to\Zc$}
(we follow the common convention of indicating ``conjugate'' indices by a dot).
We extend this Hermitian structure to the quantum bundle
by setting
$$(\Bsf^{p\a})^\#:=\Xsf_p\tn(\bb^\a)^\#
\equiv h^{\a\.\,\a}\,\Xsf_p\tn\bar\bb_{\a\.\,}\in\DCh(\Pm,\Zc)~.$$
Recalling the generalised index summation convention,
and using the spatial momentum as a generalised index,
we can see this Hermitian structure as associated with the tensor
having the generalised components
\hbox{$H^{p\.\,\a\.\,,\a p}\equiv H^{p\.\,p}\,h^{\a\.\,\a}$}
where \hbox{$H^{p\.\,p}=\d(p_\sbo{}\!\!\.\,-p_\sbo)$}\,.

\subsection{Detectors and quantum configuration space}
\label{ss:Detectors and quantum configuration space}

We consider a timelike submanifold \hbox{$\T\subset\M$}
and call it a \emph{detector}.
A momentum space formalism for particle interactions,
in terms of generalised semi-densities,
can be exhibited as a sort of a complicated `clock'
carried by the detector.

A \emph{generalised frame of free one-particle states}
along $\T$ can be obtained as follows.
At some arbitrarily fixed event \hbox{$x_0\in\T\subset\M$}
we choose a classical frame
\hbox{$\bigl(\bb_\a\bigr)$} of the bundle \hbox{$\Z\onto(\Pm)_{x_0}$}\,,
and note that the family of generalised semi-densities
$\bigl\{\Bsf_{p\a}(x_0)\bigr\}$
is a generalised frame of \hbox{$\ZC^1_{x_0}\onto(\Pm)_{x_0}$}\,.
Then we transport $\Bsf_{p\a}$ along $\T$
by means of Fermi transport~\cite{C09a,C12a}
for the spacetime and spinor factors,\footnote{
Fermi and parallel transport coincide if the detector is inertial.}
and, for the remaining factors,
by means of parallel transport relatively to a background connection of $\Z$
which will have to be assumed
(see also~\sref{ss:Remarks about gauge fixing}).
We write
$$\Bsf_{p\a}:\T\to\DCh(\Pm,\Z)_\Tt:
t\mapsto\Bsf_{p\a}(t)=\Xsf_{p(t)}\tn\bb_\a~,$$
where \hbox{$p:\T\to\Pm:t\mapsto p(t)$} is Fermi-transported
and \hbox{$t=0$} at $x_0$\,.
This yields a trivialization
$$\DCh(\Pm,\Z)_\Tt\cong\T\times\DCh(\Pm,\Z)_{x_0}~,$$
which can be seen as determined by a suitable connection called
the \emph{free-particle connection}.
Eventually, the above arguments can be naturally extended
to multi-particle bundles and states.
When several particle types are considered,
we get a trivialization \hbox{$\VC_{\!\Tt}\cong\T\times\QC$}
of the total quantum state bundle,
where \hbox{$\QC\equiv\VC_{\!x_0}$} can be seen as the
``quantum configuration space''.
The quantum interaction,
an added term that modifies the free-field connection,
can be constructed by
assembling the classical interaction with a distinguished quantum ingredient.
By construction, the free-particle transport
preserves particle type and number,
while the interaction doesn't~\cite{C05,C12a}.

Let \hbox{$(\TO\M)_{{}_\T}\onto\T$} denote the restriction
of the tangent bundle of $\M$ to base $\T$.
We have the splitting
\hbox{$(\TO\M)_{{}_\T}=
(\TO\M)_\Tt^{\spar}\oplus(\TO\M)_\Tt^{\sbot}$}
into ``timelike'' and ``spacelike'' $g$-orthogonal subbundles.
Exponentiation determines, for each $t\in\T$,
a diffeomorphism from a neighbourhood of $0$ in $(\TO\M)_t^{\sbot}$
to a spacelike submanifold $\M\!_t\subset\M$,
and so a 3-dimensional foliation of a neighbourhood
\hbox{$\N\equiv\bigcup_{t\in\T}\M\!_t\subset\M$} of $\T$.

A \emph{tempered} generalised semi-density on $(\Pm)_t$ yields,
via Fourier transform, a generalised semi-density on
\hbox{$(\TO\M)_t^{\sbot}$}\,.
A suitable restriction then yields, via exponentiation,
a generalised semi-density on $\M\!_t$
(remember~\cite{Sc} that a distribution can be restricted to an open set).
This correspondence can be extended to $\Z$-valued semi-densities
by means of background linear connections of the various ``internal'' bundles.
We may view these auxiliary connections not as classical gauge fields
but rather as a ``mean field'' background structure,
analogous to the gravitational background,
whereas quantum gauge fields are of a different nature
(more about that later).
Eventually, the trivialisation \hbox{$\VC_{\!\Tt}\cong\T\times\QC$}
can be extended as \hbox{$\VC_{\!\sst\N}\cong\N\times\QC$}\,.
Note how this essentially amounts to a natural extension
of the generalised quantum frames $\bigl(\Bsf_{p,\a}\bigr)$ over $\T$
to generalised frames over $\N$.
In flat spacetime, and for an inertial detector,
we essentially get the usual correspondence between
momentum-space and position-space representation.

Accordingly, we also get the operator algebra
$$\OC\cong\QC\tn\QC^*\equiv\VC_{\!x_0}\tn\VC_{\!x_0}^*~,$$
where, as previously discussed,
the identification is determined via normal ordering.

\subsection{Quantum fields}
\label{ss:Quantum fields}

Taking the fiber Hermitian structure of \hbox{$\Z\onto\Pm$} into account,
any \hbox{$\z\in\DCh(\Pm,\Z^*)$} yields an emission operator
\hbox{$\cre{a}[\z]\equiv\tra{a}[\z^\#]:\psi\mapsto\z^\#\swe\psi$}
and an absorption operator \hbox{$\abs{a}[\z]:\psi\mapsto\z\,|\,\psi$}\,.
We write
$$\abs{a}^{\a}(p_\sbo)\equiv\abs{a}^{p\,\a}:=\abs{a}[\Bsf^{p\a}]~,\quad
\cre{a}{}^{\a}(p_\sbo)\equiv\cre{a}{}^{p\a}:=
\tra{a}[(\Bsf^{p\a})^\#]~,$$
thus seeing $\abs{a}^{\a}$ and $\cre{a}{}^{\a}$
as generalised functions of momentum.
Consistently with the generalised index notation we also write
\hbox{$\abs{a}[\z]=\z_{p\,\a}\,\abs{a}^{p\,\a}$},
\hbox{$\cre{a}[\z]=\z_{p\,\a}\,\cre{a}{}^{p\,\a}$},
and eventually
$$\abs{a}[\_]=\abs{a}^{p\,\a}\,\Bsf_{p\,\a}~,\quad
\cre{a}[\_]=\cre{a}{}^{p\,\a}\,\Bsf_{p\,\a}~.$$

We now consider a notion of quantum field on curved spacetime
based on the choice of a detector and the related constructions
presented in~\sref{ss:Detectors and quantum configuration space}.
If $(\M,g)$ is Minkowski spacetime and $\T$ is inertial (a straight line)
then we have the orthogonal decomposition \hbox{$\M=\T\,{\times}\,\X$},
namely \hbox{$\M\!_t\equiv\X$} (a Euclidean space) \hbox{$\forall\,t\in\T$},
and define the \emph{free quantum field}, in the considered sector,
to be \hbox{$\phi(x)\equiv\phi^{\sst(+)}(x)+\phi^{\sst(-)}(x)$} with
\begin{align*}
&\phi^{\sst(+)}(x):=\frac1{(2\pi)^{3/2}}\int\frac{\dO^3 p}{\sqrt{2\,p_0}}\,
\eO^{-\iO\,\bang{p,x}}\,\abs{a}^\a(p_\sbo)\,\bb_\a(p_\sbo)~,
\\[6pt]
&\phi^{\sst(-)}(x):=\frac1{(2\pi)^{3/2}}\int\frac{\dO^3 p}{\sqrt{2\,p_0}}\,
\eO^{\iO\,p_0\,t}\,\cre{a}{}^{\a}(p_\sbo)\,\bar\bb_\a(p_\sbo)~,
\end{align*}
where \hbox{$p_0\equiv(m^2+p_\sbo^2)^{1/2}$},
\hbox{$x\equiv(t,x_\sbo)$}\,,
\hbox{$\bang{p,x}=p_0\,t+\bang{p_\sbo,x_\sbo}$}\,.
Thus $\phi^{\sst(+)}(x)$ and $\phi^{\sst(-)}(x)$ are, respectively,
a Fourier transform of $\abs{a}[\_]$
and a Fourier anti-transform of $\cre{a}[\_]$
related to the Leray form of the mass shell.\footnote{
As indicated by the factor $(2\,p_0)^{-1/2}$,
which could be absorbed either in
$\bigl\{\Bsf_{p\,\a}\bigr\}$ or in the interaction.}
In curved spacetime, all this still makes sense
if the neighbourhood $\N$ of $\T$ coincides with $\M$
(namely we have a global time$\,{\times}\,$space decomposition of $\M$).
Alternatively, it could be viewed as a kind of \emph{linearization}
by replacing \hbox{$\M\!_t\equiv\X$} with \hbox{$(\TO\M)_{x_0}^{\sbot}$}
in the whole construction.

\remark~$\abs{a}[\_]$ and $\cre{a}[\_]$ can be seen as generalised maps
\hbox{$\T\,{\times}\,(\P\!\!_\sbot\,{\times}\,\Z^*)_{x_0}\to\OC$}\,,
or equivalently, because of linearity, as morphisms
\hbox{$\T\,{\times}\,\P\!\!_\sbot \to\Z_{\!\Tt}\tn\OC$}
over $\T$ and distributions on \hbox{$\P\!\!_\sbot\onto\T$}.\smallbreak

Our next step will be expressing the above $\phi$
as a local section of some vector bundle over $\M$,
in order to recover the usual approach
in which one ``quantises'' a classical field
by rendering it ``operator-valued''.
The basic issue here is that $\Z$, in general,
is a vector bundle over $\Pm$\,.
In many practical cases one actually has a ``semi-trivial'' bundle
\hbox{$\Pm\cart{\M}\Z$}, where \hbox{$\Z\onto\M$} is the true
``internal'' vector bundle.
The notable exceptions are given by spin bundles.
For fermions, these are sub-bundles of semi-trivial bundles
anyway~(\sref{ss:Geometric background for gauge field theory}).
For gauge bosons, the same holds true when a gauge
has been chosen~(\sref{ss:Remarks about gauge fixing}).
So \hbox{$\bb_\b(p_\sbo)=\Ksf^{\a}_\b(p_\sbo)\,\bb'_\a$}
and \hbox{$\bar\bb_\b(p_\sbo)=\Ksf^{\a}_\b(p_\sbo)\,\bar\bb'_\a$}\,,
where $\bigl(\bb'_\a\bigr)$ is a new frame, independent of momentum.
The Hermitian metric yields an identification \hbox{$\smash{\Z\cong\Za}$},
allowing us to write \hbox{$\bb'_\a=\bar\bb'_\a$} and
\hbox{$\phi(x)=\phi^\a(x)\,\bb'_\a(t)$}\,, with
$$\phi^\a(x)\equiv
\frac1{(2\pi)^{3/2}}\int\frac{\dO^3 p}{\sqrt{2\,p_0}}\,
\Ksf^{\a}_\b(p_\sbo)\Bigl(
\eO^{-\iO\,\bang{p,x}}\,\abs{a}^\b(p_\sbo)
+\eO^{\iO\,\bang{p,x}}\,\cre{a}{}^\b(p_\sbo)\Bigr)~.$$
Eventually we can identify $\phi(x)$
with an element in the fiber at $x$ of a vector bundle over $\M$
by parallel transport of $\bb'_\a$ along the spacelike geodesic from $t$ to $x$
(relatively to the gauge-fixing connection).
Its components $\phi^\a(x)$ are valued in a \emph{fixed} algebra
of linear operators on the space $\QC$ of quantum states.\footnote{
We remark that $\phi$ is a combination of particle absorption
and \emph{anti-particle emission},
while the conjugate field $\bar\phi$ is a combination
of anti-particle absorption and particle emission.}

\remark~The meaning of the above integral should be specified.
The integrand can be seen as a generalised map
\hbox{$\Phi_x:\P\!\!_\sbot\to\OC^1$},
the target being the fixed vector space
of linear maps \hbox{$\QC_{\circ}\to\QC$}
introduced in~\sref{ss:Quantum states as generalised semi-densities}
(\hbox{$\QC_{\circ}\subset\QC$} is the subspace of all test elements).
Then the requirement
$$\bigl\langle \z\,,\,
\bigl(\int\!\dO^3 p\,\Phi_x(p_\sbo)\bigr)\chi\bigr\rangle=
\int\!\dO^3 p\,\bigl\langle \z\,,\,\Phi_x(p_\sbo)\chi \bigr\rangle~,
\qquad \z,\chi\in\QC_{\circ}~,$$
characterises the integral as belonging
to the extended operator space $\OC^\bullet$
(\sref{ss:Quantum states as generalised semi-densities}).
Similar observations can be made in other situations where integrals of
$\OC$-valued maps occur
(\sref{ss:Basic forms and Lagrangian field theory},\;%
\sref{ss:Symmetry and charge}).\smallbreak

Henceforth, for notational simplicity, we'll indicate as $\OC$
the above said extended operator space.
We can now see the quantum field as a generalised section
(in the distributional sense) \hbox{$\M\to\OC\ten{\M}\Z$}\,.
The set of all quantum fields of a theory
can be described as a generalised section
\hbox{$\M\to\EC\equiv\OC\tn\E$},
where \hbox{$\E\onto\M$} is the classical ``configuration bundle''
(the finite-dimensional vector bundle
whose sections are the classical fields).
In other terms, the ``quantum bundle'' \hbox{$\EC\onto\M$}
replaces the classical configuration bundle.
Note that all tensor products and contractions,
via multiplication in $\OC$,
again generate $\OC$-valued fields.
In particular, a generalised map \hbox{$\M\to\OC$}
may be called a \emph{quantum scalar}.
Finally, point-wise multiplication of field components
is super-commutative by virtue of the
modified rule~\eqref{equation:modifiedsupercommutator}.

The above $\phi$ is an essentially unique, well-defined object,
fulfilling the Klein-Gordon equation
and determined by the underlying classical geometry.
We call it a \emph{free field}.
We'll be interested, more generally, in ``interpolating fields'',
namely generalised sections \hbox{$\M\to\EC$} which are solutions
of a differential equation containing interactions
among various ``sectors'' in $\EC$\,.

\remark~Admittedly, the above scheme relies on the choice of a detector and,
in general, only makes sense as a local linearisation.
However, some kind of an observer and the preferred time related to it
are always needed in quantum physics,
and we are lead to conclude that the equivalence
between position and momentum representations may be only local and partial
in non-flat spacetime.
We might argue that the particle/field complementarity issue
is still unresolved:
the two views can't be both fundamental and on the same footing.

\section{F-smooth geometry and Lagrangian field theory}
\label{s:F-smooth geometry and Lagrangian field theory}
\subsection{F-smooth geometry for quantum fields}
\label{ss:F-smooth geometry for quantum fields}

The notion of F-smoothness enters our approach to quantum fields
essentially in two ways,
the first being the construction of bundles of states over momenta
(\sref{ss:Froelicher-smooth quantum state bundles}).
Then, in order to deal with quantum fields
as sketched in~\sref{ss:Quantum fields},
we must extend the fundamental differential geometric notions for
classical bundles to the case of a quantum bundle
\hbox{$\EC\onto\M$}.

We start from the F-smooth structure of $\OC$
characterised by a suitable set $\CCal_{\sst\!\OC}$\,,
defined to be the family of all curves
\hbox{$C:\RR\to\OC$} such that, for any test element
\hbox{$\chi\in\QC_{\circ}\subset\QC$}\,,
the curve \hbox{$C\chi:\RR\to\QC:s\mapsto C(s)\chi$} is F-smooth.
In particular, if \hbox{$\a:\RR\to\QC^{{*}1}$}
and \hbox{$c:\RR\to\QC^1$} are F-smooth curves,
then the curves
\hbox{$s\mapsto\abs{a}[\a(s)]$} and \hbox{$s\mapsto\tra{a}[c(s)]$}
are F-smooth.
A curve \hbox{$\hat C:\RR\to\OC\tn\E$} is said to be F-smooth if
$$\bang{\s,\hat C}:\RR\to\OC:s\mapsto\bang{\s,\hat C(s)}$$
is F-smooth for any classical smooth section \hbox{$\s:\M\to\E^*$}.

Now \hbox{$\wp:\EC\equiv\OC\tn\E\onto\M$}
turns out to be an F-smooth vector bundle:
if \hbox{$\X\subset\M$} is an open subset and
\hbox{$(x,y):\E_\Xx\to\X\,{\times}\,\Y$} is a smooth linear trivialisation,
then one has an F-smooth
local trivialisation \hbox{$\EC\!_\Xx\to\X\,{\times}\,\YC$},
still denoted as $(x,y)$\,,
where \hbox{$\YC\equiv\OC\tn\Y$}.
In particular, a linear fibered chart $\bigl(x^a,y^i\bigr)$
on $\E$ can be seen as a linear fibered chart on $\EC$,
the fiber ``coordinates'' $y^i$ being now $\OC$-valued.

To the reader which is familiar with the geometry of jet bundles,
parts of this section will look, at first sight, just as remainders;
this is exactly one of the virtues of the F-smoothness approach.
However we note that certain points do require some care.

\subsubsection*{Tangent, vertical and jet spaces}
It can be shown~\cite{C04a} that if $c$ is an F-smooth curve valued
into a distributional space then there exists a unique curve $\de c$,
valued into the same space, such that
\hbox{$\bang{c,u}'=\bang{\de c,u}$} for any test element $u$
(here the prime denotes the ordinary derivative
of functions \hbox{$\RR\to\RR$}).
Hence, if \hbox{$C:\RR\to\OC$} is an F-smooth curve then the rule
\hbox{$\de C(s)\chi:=\de(C(s)\chi)~\forall\chi\in\QC_{\circ}$}
defines a unique curve \hbox{$\de C:\RR\to\OC$};
moreover, the tangent prolongation
\hbox{$\TO C:\RR\times\RR\to\TO\OC\equiv\OC\,{\times}\,\OC$}
is well defined as
\hbox{$\TO C(s,\t):=\bigl(C(s),\,\t\,\de C(s)\bigr)$}.

If \hbox{$\hat C:\RR\to\EC\equiv\OC\tn\E$} is F-smooth then
\hbox{$x^a\comp\hat C:\RR\to\RR$}
and \hbox{$y^i\comp\hat C:\RR\to\OC$}.
Two such curves $\hat C$ and $\hat C_{{}_{\!1}}$
are said to be \emph{first-order equivalent} at (say) \hbox{$0\in\RR$}
if \hbox{$\hat C_{{}_{\!1}}(0)=\hat C(0)\equiv y\in\EC$}
and their ``components'' have the same derivatives at $0$\,, namely
\hbox{$\de(x^a\comp\hat C)(0)=\de(x^a\comp\hat C_{{}_{\!1}})(0)$}
and \hbox{$\de(y^i\comp\hat C)(0)=\de(y^i\comp\hat C_{{}_{\!1}})(0)$}
in any chart.
The equivalence class $\TO_{\!y}\EC$ is called
the \emph{tangent space of $\EC$ at $y$},
and naturally turns out to be a vector space.
Each local linear trivialization
\hbox{$(x,y):\E_\Xx\to\X\,{\times}\,\Y$}
determines a local F-smooth trivialization
$$\TO(x,y)\equiv(\TO x,\TO y):\TO\EC\equiv
\bigsqcup_{y\in\EC}\TO_{\!y}\EC\to\TO\X\,{\times}\,\TO\YC~,$$
where \hbox{$\YC\equiv\OC\tn\Y$}, \hbox{$\TO\YC\equiv\YC\,{\times}\,\YC$}.
Hence, the fibered set \hbox{$\pi_\EC:\TO\EC\onto\EC$}
turns out to be an F-smooth bundle.
We have another F-smooth bundle with the same total space, namely
$$\TO\wp:\TO\EC\onto\TO\M:\de\hat C\mapsto\de(\wp\comp\hat C)~.$$
Moreover we have the \emph{vertical subbundle}
\hbox{$\VO\EC:=\Ker \TO\wp\subset \TO\EC$}\,,
the natural identification \hbox{$\VO\EC=\EC\cart{\M}\EC$},
and the exact sequence over $\EC$
$$0\to\VO\EC\to \TO\EC\to \EC\cart{\M}\TO\M\to0~.$$

The subbundle of \hbox{$\TS\M\otimes_{\EC}\TO\EC$} which projects over
the identity of $\TO\M$ is called the \emph{first jet bundle}
and denoted by \hbox{$\JO\EC\to\EC$}.
This is an affine bundle over $\EC$,
with `derived' vector bundle
\hbox{$\TS\M\operatorname*{\otimes}_{\EC}\VO\EC$}.
The restriction of $\TO^*x\tn\TO(x,y)$
is a local bundle trivialization
$$\JO(x,y):\JO\EC\!_\Xx\to \JO(\X\times\YC)\cong
\YC\times\bigl(\TS\X\tn\YC\bigr)~.$$
If $\bigl(x^a,y^i\bigr)$ is a linear fibered coordinate chart of $\EC$,
then the naturally induced charts on $\TO\EC$ and $\JO\EC$
are denoted by $\bigl(x^a,y^i;\dot x^a,\dot y^i\bigr)$
and $\bigl(x^a,y^i,y^i_a\bigr)$, respectively.
Also higher jet spaces can be defined similarly to the finite-dimensional case, and have the same basic formal properties.

As the classical bundle \hbox{$\E\onto\M$} is a vector bundle,
\hbox{$\JO\E\onto\M$} turns out to be a vector bundle, too
(while \hbox{$\JO\E\onto\E$} is affine).
Then it's not difficult to exhibit a distinguished isomorphism
\hbox{$\JO\EC\cong\OC\tn\JO\E$}.
Similarly, \hbox{$\VO\EC\cong\OC\tn\VO\E$}
and \hbox{$\TO\EC\cong\OC\tn\TO\E$},
where the latter tensor product is over $\TO\M$.

\subsubsection*{Prolongations}

If \hbox{$f:\EC\to\OC$} is F-smooth then its \emph{tangent prolongation}
\hbox{$\TO f\equiv(f,\dO f):\TO\EC\to\TO\OC\equiv\OC\,{\times}\,\OC$}
is characterised by the requirement that the rule
\hbox{$\bang{\dO f,\de\hat C}=\de(f\comp \hat C)$}
holds for any F-smooth curve \hbox{$\hat C:\RR\to\EC$}.
Then \hbox{$\dO f:\TO\EC\to\OC$}
turns out to be a linear morphism
(closely related to the usual notion of \emph{functional derivative}).

Setting \hbox{$v\equiv\de\hat C(0)\in\TO\EC$}
we also write \hbox{$v.f\equiv\de(f\comp \hat C)(0)$}\,;
more generally, if \hbox{$v:\EC\to\TO\EC$} is F-smooth
then we obtain the F-smooth function \hbox{$v.f\equiv\bang{\dO f,v}$}\,,
called the \emph{derivative of $f$ along $v$}.
It's not difficult to see
that this operation fulfills the ordinary Leibnitz rule
$$v.(fg)=(v.f)\,g+f\,v.g=\bang{\dO f,v}\,g+f\,\bang{\dO g,v}~.$$
We then write \hbox{$\dO(fg)=\dO f\,g+f\,\dO g$}
with the understanding that it is applied to $v$
by applying $\dO f$ and $\dO g$ to $v$
\emph{without changing the order of any factors}.

If \hbox{$\Phi:\EC\to\EC'$}
is a morphism of quantum bundles then we define
\hbox{$\TO\Phi:\TO\EC\to\TO\EC'$} to be characterised by
\hbox{$\TO\Phi\comp\de\hat C=\de(\Phi\comp\hat C)$},
and \hbox{$\VO\Phi:\VO\EC\to\VO\EC'$} as its vertical restriction.

If \hbox{$\phi:\M\to\EC$} is an F-smooth section,
then $\TO\phi:\TO\M\to\TO\EC$ projects over the identity of $\TO\M$,
so that it can be viewed as a section \hbox{$\jO\phi:\M\to\JO\EC$}
(the \emph{first jet prolongation} of $\phi$).
The \emph{jet prolongation morphism} \hbox{$\JO\Phi:\JO\EC\to\JO\EC'$}
is defined by the requirement \hbox{$\jO(\Phi\comp\phi)=\JO\Phi\comp\jO\phi$}
for any $\phi$\,.
We then get the tangent, vertical and jet functors,
which can be iterated essentially as in finite-dimensional geometry.
In particular, we indicate the $k$-th (holonomic) jet prolongation of
\hbox{$\EC\onto\M$} as
\hbox{$\JO_k\EC\onto\JO_{k-1}\EC\onto\mdots\onto\EC\onto\M$}.

The induced fiber coordinates $y^i_\sA$ on jet bundles,
where $\scriptstyle{A}$ is a base manifold multi-index,
are defined as usual by
\hbox{$y^i_\sA\comp\jO_k\phi=\de_\sA\phi^i$}
for all sections \hbox{$\phi:\M\to\EC$},
\hbox{$k=|{\scriptstyle{A}}|$} being the multi-index' length.

Straightforward extensions of the above constructions
yield bundles $\VO\JO_k\EC$ and $\JO_k\VO\EC$,
which turn out to be naturally isomorphic to each other.

\subsubsection*{Partial derivatives}

Let \hbox{$\bigl(x^a,y^i\bigr)$} be a linear fibered coordinate chart of
\hbox{$\E\onto\M$}.
The fiber coordinates $\bigl(y^i\bigr)$ can be seen as elements of
the dual frame of a local frame $\bigl(\bb_i\bigr)$\,,
and as $\OC$-valued fiber coordinates on $\EC$.
If \hbox{$f:\EC\to\OC$} then the partial derivatives $\dde{}{x^a}f$
are well defined,
since the coordinate curves \hbox{$\RR\to\M$}
corresponding to the base coordinates $\bigl(x^a\bigr)$
can be seen as $\EC$-valued
via the trivialisation associated with the chart itself.

Instead, the coordinate curves corresponding
to the fiber coordinates $\bigl(y^i\bigr)$
can't be naturally seen as $\EC$-valued,
so that we have no obvious meaning for \hbox{$\de_if\equiv\de f/\de y^i$},
and wonder whether we can apply the known results and coordinate formulas
of classical Lagrangian field theory on jet bundles.
Now observe that a vertical vector field \hbox{$v:\EC\to\VO\EC$}
can be written as $v^i\de_i$\,,
where \hbox{$\de_i\equiv{\smash{\dde{}{y^i}}}$} denotes the classical vectors
tangent to the fiber coordinate curves
and the components $v^i$ are $\OC$-valued.
Moreover \hbox{$\bang{\dO f,v}=v.f$} has a well-defined meaning,
as previously discussed.
In particular, if \hbox{$\l\in\OC$} then we write
\hbox{$\bang{\dO f,\l\,\de_i}\equiv\de_if\l\equiv\de_if(\l)$}\,,
namely we regard $\de_if$ as a function on $\EC$
\emph{valued into linear maps} \hbox{$\OC\to\OC$}.
Thus \hbox{$\bang{\dO f,v}=\de_if\,v^i$} gets a well-defined meaning,
though the right-hand side must \emph{not} be intended
as ordinary multiplication.
We now write \hbox{$\dO f=\de_af\,\dO x^a+\de_i f\,\dO y^i$},
so that if \hbox{$v=v^a\,\de_a+v^i\,\de_i:\EC\to\TO\EC$}
is an arbitrary vector field then
\hbox{$\bang{\dO f,v}=\de_af\,v^a+\de_i f\,v^i$}.

In practical calculations one may wish to be able to write down
an explicit expression for $\de_if$\,.
This can be easily done when $f$ is a polynomial
in the linear fiber coordinates $y^i$,
as it is indeed true in many cases of interest (\sref{ss:The Lagrangian}).
One possibility would be marking in some way
the ``insertion point'' for the $\OC$-argument in each monomial in $\de_if$\,.
But recalling (\sref{ss:Quantum fields})
that the product of field components at any given spacetime point
is defined in such a way to be super-commutative,
and observing that the fiber coordinates
can be (and are) chosen in such a way that
each one belongs to a sector of definite $\ZZ_2$-grade,
we can just define \hbox{$\dde{f}{y^i}\equiv\de_if$} by
$$\dde{f}{y^i}\,\l=\bang{\dO f,\l\,\de_i}\quad
\forall\l\in\OC~\text{such that}~\grade{\l}=\grade{y^i}~.$$
In this way, explict expressions of partial derivatives
are the same as in the classical situation,
\emph{possibly up to a sign in each monomial}.

The above discussion naturally extends to partial derivation
with respect to the induced fiber coordinates $y^i_\sA$ on jet bundles.
As a consequence, several basic differential geometric operations
on the ``quantum bundle'' \hbox{$\EC\onto\M$}
turn out to have, formally, the same expressions
as their classical counterparts
on the ``classical bundle'' \hbox{$\E\onto\M$},
though some extra care will be needed in the ordering of factors.

\subsubsection*{Connections}

We denote by \hbox{$\vartheta:\JO\EC\cart{\EC}\TO\EC\to\VO\EC$}
the complementary morphism, over $\EC$, of the inclusion
\hbox{$\JO\EC\into\TS\M\otimes_{\EC}\TO\EC$}.
Its coordinate expression is
\hbox{$\vartheta^i=\dO y^i-y_a^i\,\dO x^a$}\,.
The basic idea associated with $\vartheta$ is that
a point \hbox{$\xi\in\JO_e\EC$}\,, \hbox{$e\in\EC$},
determines a linear projection \hbox{$\TO\!_e\EC\onto\VO\!_e\EC$}
and hence a splitting of $\TO\!_e\EC$ into the direct sum
\hbox{$\HO_\xi\EC\oplus\VO\!_e\EC$} of a ``$\xi$-horizontal'' subspace
and the vertical subspace.

A \emph{connection} on \hbox{$\wp:\EC\to\M$}
is defined to be an F-smooth section \hbox{$\G:\EC\to\JO\EC$}.
As in the finite-dimensional case,
a connection can be assigned by choosing any one
of various equivalent structures.
First, $\G$ can be viewed as a linear morphism
\hbox{$\EC\cart{\M}\TO\M\to\TO\EC$} over $\EC$
such that \hbox{$(\pi_\EC\,,\TO\wp)\comp\G$}
turns out to be the identity of \hbox{$\EC\cart{\M}\TO\M$}.
The image
\hbox{$\HO_{\sst\G}\EC:=\G(\EC\cart{\M}\TO\M)$}
is a vector subbundle of \hbox{$\TO\EC\onto\EC$}\,;
the restriction of \hbox{$\G\comp(\pi_\EC\,,\TO\wp)$}
is the identity of $\HO_{\sst\G}\EC$.
If \hbox{$v:\M\to\TO\M$} is a smooth vector field,
then $\G[v]:\EC\to\TO\EC$ is an F-smooth vector field,
called its \emph{horizontal lift}.
Moreover we have the complementary morphism
$$\Om_\ssGa:=
\vartheta\comp\G\equiv\id-\G:\TO\EC\to\VO\EC\equiv\EC\cart{\M}\EC~,$$
so that the map \hbox{$\bigl(\G\comp(\pi_\EC\,,\TO\wp)\,,\,\Om_\ssGa\bigr)$}
determines the decomposition
\hbox{$\TO\EC=\HO_{\sst\G}\EC\oplus_{\EC}\VO\EC$}\,.

The \emph{covariant derivative} of an F-smooth section \hbox{$\phi:\M\to\EC$}
is defined to be the linear morphism over $\M$
$$\nabla\phi\equiv\nabla[\G]\phi:=
\pr2\comp\,\Om_{\sst\G}\comp\TO\phi:\TO\M\to\EC~.$$
If \hbox{$v:\M\to\TO\M$} is a  vector field then we write
\hbox{$\na_v\phi\equiv\nabla\phi\comp v$}.

We say that $\G$ is a \emph{linear connection} if it is a linear morphism
\hbox{$\EC\to\JO\EC$} over $\M$.
Then its \emph{curvature tensor} can be defined as a section
\hbox{$\Rcal_\ssGa:\M\to\weu2\TS\M\ten{\M}\End(\EC)$}\,,
formally characterised  as in the finite-dimensional case.

Any linear morphism of finite-dimensional vector bundles
naturally yields a linear morphism of the corresponding quantum bundles
(via tensor product by $\OC$).
In particular, a linear connection \hbox{$\G:\E\to\JO\E$}
of the classical configuration bundle yields a linear quantum connection
\hbox{$\EC\to\JO\EC$},
which is indicated by the same symbol.

\subsubsection*{Dual quantum bundles}

The cotangent space is one notion of classical geometry
which does not readily extends to the context of F-smooth geometry.
Such extension can be done in the present special context,
still without being involved with duality in a general sense.

We define the \emph{dual bundle} of \hbox{$\EC\equiv\OC\tn\E\onto\M$}
to be simply \hbox{$\EC^*\equiv\OC\tn\E^*\onto\M$},
so avoiding functional duals.
An element in \hbox{$\EC^*_{\!x}$}\,, \hbox{$x\in\M$},
can be seen as a linear map \hbox{$\EC_{\!x}\to\OC$}.
Moreover we define the \emph{vertical dual}
to be \hbox{$\VS\EC:=\EC\cart{\M}\EC^*$}.
If \hbox{$f:\EC\to\OC$} then its \emph{fiber differential}
\hbox{$\check\dO f:\EC\to\VS\EC$} is well-defined as
the restriction of $\dO f$ to $\VO\EC$.

A straightforward extension of the above procedure
yields the bundles $\VO^*\JO_k\EC$, \hbox{$k\in\NN$}\,.

In order to introduce the \emph{cotangent bundle}
\hbox{$\TS\EC\onto\EC$} we first note that any
\hbox{$\l\in\OC\tn\TS\M$} can be viewed as an F-smooth linear 
morphism \hbox{$\TO\EC\onto\OC$} over $\EC$
via the rule \hbox{$v\mapsto\bang{\l,\TO\wp(v)}$}.
The image of this inclusion is a vector bundle \hbox{$\HO^*\EC\onto\EC$}.
We define \hbox{$\TS\EC\onto\EC$} to be be the smallest vector bundle,
containing $\HO^*\EC$ as a sub-bundle,
whose fibers are constituted by F-smooth linear 
maps \hbox{$\TO\EC\onto\OC$} whose restrictions to $\VO\EC$
are in $\VS\EC$.
As in the classical case we have the exact sequence over
\hbox{$0 \to \HO^*\EC \to \TS\EC \to \VS\EC \to 0$}\,,
which splits over \hbox{$\JO\EC$}.

\subsection{Basic forms and Lagrangian field theory}
\label{ss:Basic forms and Lagrangian field theory}

In classical Lagrangian field theory~%
\cite{MaMo83b,Kolar84,KrupkaKrupkovaSanders10,KrupkaSanders08,
VinogradovCSS84I,VinogradovCSS84II}
one deals with arbitrary exterior forms
on jet bundles of the configuration bundle.
Furthermore one deals with the Lie bracket of arbitrary vector fields
and the Fr\"olicher-Nijenhuis bracket of vector-valued forms.
All such notions and operations can be introduced
in the above described quantum context,
though some complications do arise.
For our present purposes, however, we'll only need the restricted notions
of vertical dual and of
\emph{basic} (or \emph{totally horizontal}) \emph{form}.

A ``basic'' $q$-form of order $k$ is defined to be an F-smooth morphism
\hbox{$\a:\JO_k\EC\to\OC\tn\weu{q}\TS\M$} over $\M$,
\hbox{$q,k\in\{0\}\cup\NN$}\,.
A basic $0$-form, in particular, is a map \hbox{$f:\JO_k\EC\to\OC$}.
It's easy to see that there exists a unique $1$-form of order $k\,{+}\,1$\,,
indicated as \hbox{$\dH f:\JO_{k+1}\EC\to\OC\tn\TS\M$}
and called the \emph{horizontal differential} of $f$,
with the property that for any F-smooth section \hbox{$\phi:\M\to\EC$} one has
\hbox{$\dH f\comp\jO_{k+1}\phi=\dO(f\comp\jO_k\phi)$}\,.
We write its coordinate expression as $\dO_af\,\dO x^a$, with
$$\dO_af\equiv
\de_af+\sum_{0\leq|\sA|\leq k}\de_i^\sA f\,y^i_{\sA+a}\equiv
\de_af+\de_if\,y^i_a+\de_i^bf\,y^i_{ab}+\mdots
+\de_i^{b_1\dots b_k}f\,y^i_{ab_1\dots b_k}~,$$
where \hbox{$\de_i^\sA f\equiv \de f/\de y^i_\sA$}\,.
Note that $\dO_a$ can be seen as the coordinate expression 
of the holonomic restriction of $\JO f$,
the jet functor applied to $f$ seen as a morphism over $\M$.
Similarly we indicate by $\dO_\sA f$ the expression 
of the holonomic restriction of $\JO\mdots\JO f$,
the jet functor iterated $|{\scriptstyle A}|$ times.

We extend $\dH$ to act on arbitrary basic forms\footnote{
For a general definition of horizontal and vertical differentials
in the finite-dimensional context see \eg\ Saunders~\cite{Sa89}.}
by requiring it to be an anti-derivation of degree $1$
vanishing on closed classical basic forms.
If \hbox{$\a=\a_{a_1\dots a_q}\,\dO x^{a_1}\we\mdots\we\dO x^{a_q}$}\,,
where \hbox{$\a_{a_1\dots a_q}:\JO_k\EC\to\OC$}\,,
then we get the coordinate expression
\hbox{$\dH\a=\dH\a_{a_1\dots a_q}\we\dO x^{a_1}\we\mdots\we\dO x^{a_q}$}\,.
It's then immediate to check that the horizontal differential is nilpotent,
\ie\ \hbox{$\dH^2=0$}\,.

\medbreak
Let \hbox{$m=\dim\M$}.
A \emph{$k$-order Lagrangian density} is an $m$-form
\hbox{$\Lcal:\JO_k\EC\to\OC\tn\weu{m}\TS\M$}.
We write its coordinate expression as \hbox{$\ell\,\dO^mx$}\,,
with \hbox{$\ell:\JO_k\EC\to\OC$}.
The related field equation for fields \hbox{$\phi:\M\to\EC$}
is \hbox{$\Fcal_i\comp\jO_{2k}\phi=0$}\,, where
$$\Fcal_i:=\sum_{0\leq|\sA|\leq k}\!\!\!\!
(-1)^{|\sA|}\,\dO_\sA\de^\sA_i\ell\equiv
\de_i\ell-\dO_a\de^a_i\ell+\dO_{ab}\de^{ab}_i\ell-\mdots$$
are the components of the \emph{Euler-Lagrange operator}
\hbox{$\Fcal[\Lcal]:\JO_{2k}\EC\to\weu{m}\TO^*\M\ten{\EC}\VO^{*\!}\EC$}.
As in the classical case,
the field equation can be derived from the requirement
that the action integral $\int\!\Lcal\comp\jO_k\phi$
be stationary on any compact set
for any variation of the field $\phi$ which is fixed on the set's boundary
(in other terms, the functional derivative of the action vanishes
whenever it is applied to an ``increment'' with compact support).

A straightforward extension of the notion of fiber differential
introduced in~\sref{ss:F-smooth geometry for quantum fields}
yields the fiber derivative of a basic form,
which for the Lagrangian density reads
$$\check\dO\Lcal=\dO^mx\tn\check\dO\ell:
\JO_2\EC\to\weu{m}\TS\M\ten{\EC}\VO^*\JO_k\EC~.$$
Introducing the shorthand \hbox{$P_i^\sA\equiv\de^\sA_i\ell$} we have
$$\Fcal[\Lcal]=\dO^mx\tn\Bigl(\sum_{0\leq|\sA|\leq k}\!\!\!\!
(-1)^{|\sA|}\,\dO_\sA P^\sA_i\Bigr)\,\dO y^i~,\qquad
\check\dO\Lcal=\dO^mx\tn\Bigl(
\sum_{0\leq|\sA|\leq k}\!\!\!\!P^\sA_i\,\dO y^i_\sA\Bigr)~.$$

\subsection{Vertical infinitesimal symmetries and currents}
\label{ss:Vertical infinitesimal symmetries and currents}

We consider a morphism \hbox{$v:\JO\EC\to\VO\EC$} over $\EC$,
written in coordinates as $v^i\,\de_i$ with \hbox{$v^i:\JO\EC\to\OC$}.
Its $k$-th holonomic prolongation
$$v_{\sst(k)}=\sum_{0\leq|\sA|\leq k}\!\!\!\!\dO_\sA v^i\,\de_i^\sA:
\JO_{k+1}\EC\to\VO\JO_k\EC$$
can be introduced as the restriction of the $k$-jet prolongation
\hbox{$\JO_kv:\JO_k\JO\EC\to\JO_k\VO\EC$}
to the holonomic subbundle \hbox{$\JO_{k+1}\EC\subset\JO_k\JO\EC$},
taking the natural isomorphism \hbox{$\JO_k\VO\EC\cong\VO\JO_k\EC$}
into account.

We now define a new operation $\d[v]$ acting on basic forms
\hbox{$\a:\JO_k\EC\to\OC\tn\weu{q}\TS\M$} as
$$\d[v]\a:=\check\dO\a\pint v_{\sst(k)}:\JO_{k+1}\EC\to\OC\tn\weu{q}\TS\M~.$$
This is, in a sense, a generalization of the standard Lie derivative;
but note that it raises the order of the form acted on.
The extension to non-vertical vector fields and non-basic forms is possible
but more intricate,
and won't be needed in this paper.

\remark~Setting
$$\LA_k^q:={\mathrm{Sections}}(\JO_k\EC\to\OC\tn\weu{q}\TS\M)~,
\quad k,q\in\{0\}\cup\NN\,,~0\leq q\leq\dim\M\,,$$
where \hbox{$\JO_0\EC\equiv\EC$}\,,
then $\forall q$ and for any given \hbox{$v:\JO\EC\to\TO\EC$}
we obtain the (infinite) sequence
$$\begin{CD}
\LA_{0}^q @>{\d[v]}>>
\LA_{1}^q @>{\d[v]}>> \dots
@>{\d[v]}>> \LA_{k}^q
@>{\d[v]}>> \dots
\end{CD}$$

By direct calculations it is not difficult to check:
\begin{proposition}\label{prop:dHanddeltavcommute}
The operations $\dH$ and $\d[v]$ commute.
\end{proposition}

In the case of the Lagrangian density we obtain the coordinate expression
$$\d[v]\Lcal=\Bigl(
\sum_{0\leq|\sA|\leq k}\!\!\!\!P^\sA_i\,\dO_\sA v^i\Bigr)\,\dO^mx~.$$
The above horizontal form has a straightforward physical interpretation.
Actually it's not difficult to see that if \hbox{$\phi:\M\to\EC$}
and we make the replacement \hbox{$\phi\to\phi+\ep\,v\comp\jO_k\phi$}\,,
then the variation of the action functional
\hbox{$\Ical[\phi]\equiv\int\!\Lcal\comp\jO_k\phi$}\,,
at first order in \hbox{$\ep\in\RR^+$},
is just \hbox{$\ep\int\!\d[v]\Lcal\comp\jO_{k+1}\phi$}.
Accordingly, we say that $v$ is a symmetry of the field theory
under consideration if \hbox{$\d[v]\Lcal=0$}\,.

Observing that there is a natural inclusion \hbox{$\VO^*\EC\into\VO^*\JO_k\EC$}
(complementary to the fibering \hbox{$\VO\JO_k\EC\onto\VO\EC$}),
we can compare $\check\dO\Lcal$ and $\Fcal[\Lcal]$\,.
Contracting their difference with $v_{\sst(k)}$ we get
the new $2k$-order density
$$\d[v]\Lcal-\Fcal\pint v=
\sum_{1\leq|\sA|\leq k}\Bigl(
P^\sA_i\,\dO_\sA v^i-(-1)^{|\sA|}\,\dO_\sA P^\sA_i\,v^i\Bigr)\,\dO^mx~.$$

\begin{theorem}\label{th:splitting_dvL}
$\d[v]\Lcal-\Fcal\pint v$ is an exact horizontal differential.
\end{theorem}
\proof~Consider
\hbox{$\Pcal=\dO x_a\tn\Pcal^a:
\JO_{2k-1}\EC\to\weu{m-1}\TS\M\tn\VO^*\JO_{k-1}\EC$}\,,
where \hbox{$\dO x_a\equiv\de_a|\dO^mx$} and
$$\Pcal^a=\sum_{\substack{0\leq|\sB|\leq k-1 \\ 0\leq|\sA|\leq k-|\sB|-1}}
(-1)^{|\sA|}\,\dO_\sA P_i^{a+\sA+\sB}\,\dO y_\sB^i~.$$
Then one may check that actually
\hbox{$\dH(\Pcal\pint v_{\sst(k-1)})=\d[v]\Lcal-\Fcal\pint v$}.\qed

An analogous of $\Pcal$, or an exterior form related to it,
is called  ``momentum'' in classical Lagrangian field theories,
where it is related to the notion
of \emph{Poincar\'e-Cartan form}~\cite{Kolar84,KrupkaKrupkovaSanders10}.
We'll be only involved with orders one and two,
respectively yielding
$$\Pcal^a\pint v=P^a_i\,v^i\qquad\text{and}\qquad
\Pcal^a\pint v_{\sst(1)}=(P^a_i-\dO_b P^{ab}_i)\,v^i+P^{ab}_i\,\dO_b v^i~.$$

\begin{definition}
A horizontal form \hbox{$\Jcal:\JO_{r}\EC\to\OC\tn\weu{m-1}\TS\M$}
is called a \emph{conserved current}
if \hbox{$\dH\Jcal\comp\jO_{r+1}\phi$}
vanishes for any critical section \hbox{$\phi:\M\to\EC$}
(that is, for any section fulfilling the field equation
\hbox{$\Fcal_i\comp\jO_{2k}\phi=0$}).\end{definition}

Using theorem~\ref{th:splitting_dvL} we then immediately prove
the following version of \emph{Noether's theorem}:
\begin{theorem}\label{th:Noether}
Let \hbox{$v:\JO\EC\to\VO\EC$} be a morphism over $\EC$
and \hbox{$\Ncal:\JO_k\EC\to\OC\tn\weu{m-1}\TS\M$} a basic form
such that \hbox{$\d[v]\Lcal=\dH\Ncal$}.
Then \hbox{$\Jcal[v]:=\Pcal\pint v_{\sst(k-1)}-\Ncal$}
is a conserved current.
\end{theorem}

\subsection{Symmetry and charge}
\label{ss:Symmetry and charge}

In relation to the setting discussed in~\sref{ss:Quantum fields}
we now consider a spacetime synchronization,
and use coordinates adapted to it:
synchronicity submanifolds are characterized by \hbox{$x^0=\text{constant}$}.
We set \hbox{$\p_i\equiv P^0_i\equiv\de^0_i\ell$}\,.
In a Hamiltonian setting,
which we are not discussing in detail here,
this plays the role of the ``conjugate momentum'' associated with $y^i$.
Field components and canonical momenta fulfill
the \emph{equal-time super-commutation rules}
\begin{align*}
&\Suc{\phi^i(x)\,,\,\p_j(x')}=\iO\,\d^i_j\,\d(x_\sbo\,{-}\,x'_\sbo)\,\rdg(x)~,
\\[6pt]
&\Suc{\phi^i(x)\,,\,\phi^j(x')}=\Suc{\p_i(x)\,,\,\p_j(x')}=0~,
\end{align*}
where \hbox{$x\equiv(t,x_\sbo)$}\,, \hbox{$x'\equiv(t,x'_\sbo)$}\,,
and we used the shorthand \hbox{$\p_i(x)\equiv\p_i\comp\jO\phi(x)$}\,.
Indeed, these rules can be directly checked to hold true for free fields,
and their validity for critical sections can be inferred by arguments
based on the form of the dynamics.
Note that the product of field components valued at different spacetime points
is defined in terms of the non-super-commutative product in $\OC$
(\sref{ss:Finitely-generated multi-particle algebra}),
and related formulas must be intended in a generalized
distributional sense.

If \hbox{$\Jcal=\Jcal^a\,\dO x_a:\JO\EC\to\OC\tn\weu3\TS\M$}
is a conserved current then
$$\Jcal^0=\p_i\,v^i:\JO\EC\to\OC$$
is called the associated \emph{charge density}.
For any critical section \hbox{$\phi:\M\to\EC$}
we define the related \emph{charge} \hbox{$\brstQ\in\OC$} as
$$\brstQ:=\int_{\Mm\!_t}\Jcal\comp\jO\phi=
\int_{\Mm\!_t}(\Jcal^0\comp\jO\phi)\,\dO x_0\equiv
\int_{\Mm\!_t}\p_i(x)\,v^i(x)\,\dO^3x~,$$
where $\M\!_t$ is any synchronicity manifold.
This is independent of $t$ because $\phi$ is critical,
and certainly finite if $\phi$ has compact spatial support.
Now we suppose \hbox{$\grade{\brstQ}=\grade{\Jcal^0}=0$}\,, so that
$$\Suc{\brstQ,\phi^i(x)}=\bigl[\brstQ,\phi^i(x)\bigr]=
\int\dO^3x'\,\bigl[\p_j(x')\,v^j(x')\,,\,\phi^i(x)\bigr]~,$$
where \hbox{$v^j(x')\equiv v^j\comp\jO\phi(x')$} and the like.
Moreover we suppose \hbox{$\Suc{\phi^i(x)\,,\,v^j[\phi](x')}=0$}\,.
Then
\begin{align*}
\bigl[\brstQ,\phi^i(x)\bigr]
&=\pm\int\dO^3x'\,\Suc{\p_j(x')\,,\,\phi^i(x)}\,v^j(x')=
-\iO\int\dO^3x'\,\d^i_j\,\d(x_\sbo-x'_\sbo)\,\rdg(x)\,v^j(x')=
\\[6pt]
&=-\iO\,v^i[\phi](x)\,\rdg(x)~.
\end{align*}
Since \hbox{$v^i[\phi]\equiv\d[v]\phi^i$},
we can rewrite the above result as
$$\d[v]\phi^i=\iO\,\bigl[\brstQ,\phi^i\bigr]\,\detg^{-1/2}~.$$
One also says that the infinitesimal symmetry $v$
is \emph{generated} by the corresponding charge $\brstQ$\,.

We recall that the \emph{functional derivative} $\DO F[\phi]$
of a ``field functional'' \hbox{$\phi\mapsto F[\phi]\in\OC$}
is the linear map acting on fields \hbox{$\th:\M\to\EC$} as
\hbox{$\DO F[\phi][\th]:=
\lim_{\ep\to0}\frac1\ep\,(F[\phi\,{+}\,\ep\,\th]-F[\phi])$}\,.
Then we set
$$\d[v]F[\phi]:=\DO F[\phi][v\comp\jO\phi]~,$$
which provides a natural extension of $\d[v]$ since,
when \hbox{$F[\phi]=F\comp\jO_k\phi$} with \hbox{$F:\JO_k\EC\to\OC$}\,,
then indeed we have \hbox{$(\d[v]F)\comp\jO_k\phi=\d[v]F[\phi]$}\,.
The relation
\hbox{$\d[v]F=\iO\,\bigl[\brstQ,F\bigr]\,\detg^{-1/2}$}
can be easily seen to hold at least when $F$ is a fiber polynomial.

\section{Gauge theory and BRST symmetry}
\label{s:Gauge theory and BRST symmetry}
\subsection{Remarks about gauge fixing}
\label{ss:Remarks about gauge fixing}

If the ``matter field'' of a classical theory
is a section of a vector bundle \hbox{$\E\onto\M$},
then the classical ``gauge field'' is a linear connection of that bundle,
whose true physical meaning is encoded in the curvature tensor.
Hence, different connections may yield the same physical field.
Locally, connections can be described as tensor fields
by choosing a gauge, namely a local ``flat'' connection $\g_0$\,:
an arbitrary linear connection $\g$ is then characterised
by the difference
$$\a\equiv\g-\g_0:
\M\to\TS\M\ten{\M}\End\E\equiv\TS\M\ten{\M}\E\ten{\M}\E^*~.$$
The fibers of the vector bundle \hbox{$\End\E\onto\M$}
are constituted by all linear endomorphisms of the respective fibers of $\E$,
and are naturally Lie algebras via the ordinary commutator.
In fact, this is the Lie algebra bundle of the group bundle
\hbox{$\Aut\E\onto\M$} of all fibered automorphisms of $\E$.
Moreover $\E$ is usually endowed
with some fibered geometric structure,
which selects the (``internal'' symmetry)
Lie-group subbundle \hbox{$\Gb\onto\M$}
of all automorphisms preserving it;
the fibers of $\Gb$ are isomorphic Lie groups,
though distinguished isomorphisms among
them don't exist in general.\footnote{
Usually people prefer to deal with a \emph{fixed group},
by exploiting the notion of principal bundle.}
A section \hbox{$\M\to\Gb$} is called a (local) \emph{gauge transformation}.
The Lie algebra bundle of $\Gb$ is a sub-bundle \hbox{$\Lie\subset\End\E$}.
If we restrict ourselves to consider connections which make
the fiber geometric structure covariantly constant,
then the difference of any two such connections is $\Lie$-valued.

A linear connection can be seen as a section \hbox{$\M\to\GA$},
where \hbox{$\GA\subset\JO\E\ten{\M}\E^*\onto\M$}
is the affine sub-bundle projecting over the identity $\Id{\E}$.
Its ``derived'' vector bundle
(the bundle of ``differences of linear connections'')
is \hbox{$\DO\GA=\TS\M\ten{\M}\End\E$}\,.
A fiber symmetry determines an affine sub-bundle
\hbox{$\GAG\subset\GA$}\,, and
\hbox{$\DO\GAG=\TS\M\ten{\M}\Lie$}\,.

Gauge freedom can be viewed in terms of momenta by observing
that one usually tries to describe
a \emph{radiative} electromagnetic field as a tensor field
of the form \hbox{$F=k\we b$}\,,
with \hbox{$k,b:\M\to\P$} such that $k^\#$ is a geodesic null vector field
and \hbox{$g^\#(k,b)=0$}\,.
Though in curved spacetime there is no guarantee that
we can find a \emph{closed} such $F$,
a natural viewpoint shift suggest that this algebraic type is suitable
for describing photons.
Then the couple $(k,b)$ constitutes a \emph{redundant} description.

Recalling the notion of a quantum field discussed
in~\sref{ss:Quantum fields}
and~\sref{ss:F-smooth geometry for quantum fields},
we realize that all fields must be sections of some vector bundle.
This is achieved by fixing a gauge,
but the resulting theory has now too many ``degrees of freedom''.
In order to view this aspect in terms of the momentum formulation,
we treat a gauge field as a section \hbox{$\a:\P\to\P\tn\Lie$}\,,
\hbox{$\P\equiv\TS\M$}.
A close examination at point interactions in terms of 2-spinors~\cite{C14a}
shows that the replacement \hbox{$\a(p)\to p\tn\chi(p)+\a(p)$}\,,
with \hbox{$\chi:\P\to\Lie$}\,,
actually does not affect scattering matrix calculations.
Hence the physical meaning of the gauge field is encoded
in its equivalence class,
$\a$ and $\a'$ being equivalent if their difference is of the kind $p\tn\chi$\,.
The equivalence class of $\a$ also uniquely determines the
``curvature-like'' tensor
$$\r[\a]:=\iO\,p\we\a+\a\bwe\a~,$$
where the notation $\a\bwe\b$
stands for exterior product of $\Lie$-valued forms together with composition
(if $\a$ and $\b$ are $\Lie$-valued 1-forms
then \hbox{$(\a\bwe\b)\iI{ab}i=\sco ijk\a\iI aj\,\b\iI bk$} where
\hbox{$\sco ijk\equiv[\lfr_j\,,\,\lfr_k]^i$}
are the ``structure constants'' in the chosen special frame
$\bigl(\lfr_i\bigr)$ of $\Lie$).
The Lagrangian density for the field $\a$\,,
written in terms of $\r[\a]$\,,
contains all the needed self-interaction terms.

In scattering matrix calculations,
gauge freedom is exploited by a suitable gauge fixing,
namely by inserting, in the gauge particle propagator,
terms that won't affect the final result.
On turn, these derive from an added term in the Lagrangian
which is not gauge-invariant,
namely does not ``pass to the quotient'' when we
deal with the above said equivalence classes,
though it is a natural geometric object when $\a$ is seen as a tensor field.

\subsection{Geometric background for gauge field theory}
\label{ss:Geometric background for gauge field theory}

In previous papers~\cite{C98,C00b,C07,C10a},
a fairly general geometric background for gauge field theories,
encompassing at least all the sectors needed in the standard model,
was shown to arise naturally from a few elementary blocks
(``minimal geometric data'').
While having that in mind,
for a preliminary discussion we can conveniently use a simplified
setting in which we provisionally disregard the differences
between the right-handed and the left-handed sectors,
and issues related to symmetry breaking.

Let \hbox{$\W\onto\M$} be the bundle of Dirac spinors,
and $m$ the considered fermion's mass.
The ``semi-trivial'' bundle \hbox{$\Pm\cart{\M}\W\onto\Pm$}
has the distinguished decomposition
\hbox{$\W^+\dir{\Pm}\W^-$} with \hbox{$\W^\pm_p:=\ker(m\mp\g_p)$}\,,
where $\g$ denotes the Dirac map.
We call $\W^+$ and $\Wc{}^-$ the \emph{electron} and \emph{positron} bundles,
respectively~\cite{C00b,C07}.
The sub-bundles $\W^\pm$ are mutually orthogonal in the Hermitian metric
associated with Dirac conjugation;
this has signature $({+}\,{+}\,{-}\,{-})$\,,
and the sign of its restriction to $\W^\pm$ is the same as the label.
We have a distinguished transformation
expressing a \emph{Dirac frame}
\hbox{$\bigl(\uu_\sA(p)\,;\,\vv_\sB(p)\bigr)$},
\hbox{${\scriptstyle A},{\scriptstyle B}=1,2$},
which is adapted to the decomposition
\hbox{$\W_{\!p}=\W^+_{\!p}\oplus\W^-_{\!p}$}\,,
in terms of a frame independent of $p$,
\eg\ the Dirac frame $\bigl(\z_\a\bigr)$
associated with the observer (\hbox{$\a=1,2,3,4$}).
This transformation is inserted in the integral defining the components
of the fermion field (\sref{ss:Quantum fields}).

Let now \hbox{$\F\onto\M$} be a complex vector bundle,
endowed with a fibered positive Hermitian metric,
describing the internal fermion structure besides spin,
and \hbox{$\Lie\subset\End\F$} the associated Lie algebra bundle over $\M$
(\sref{ss:Remarks about gauge fixing}).
Our classical configuration bundle has then a fermion sector
\hbox{$\Y\equiv\W\ten{\M}\F$}
and the gauge field sector \hbox{$\P\ten{\M}\Lie$}\,.
In order to deal with gauge symmetry we'll have to include also
some ``ghost sectors'',
whose classical configuration bundles are either $\Lie$ or $\Lie^*$.
Besides the Dirac frame,
we deal with a frame $\bigl(\ff_i\bigr)$ of $\F$
and a frame $\bigl(\lfr_\sI\bigr)$ of $\Lie$\,.
Whenever no confusion arises,
a dual frame is denoted by the same symbol and shifted index position.
We also write \hbox{$\lfr_\sI=\smash{\lfr\iIi\sI ij\,\ff_i\tn\ff^j}$}
and \hbox{$\lfr^\sI=\smash{\lfr\IiI\sI ij\,\ff^i\tn\ff_j}$}\,,
so that the natural inclusion \hbox{$\lfr:\Lie\into\End\F$}
has the local expression
\hbox{$\lfr=\smash{\lfr^\sI\tn\lfr\iIi\sI ij\,\ff_i\tn\ff^j}$}.
We also have a Hermitian structure on $\End\F$,
given by \hbox{$(X,Y)\mapsto\Tr(X^\dagger\,Y)$};
its restriction to $\Lie$\,,
constituted by all traceless anti-Hermitian endomorphisms, is negative-definite,
thus \hbox{$\smash{\lfr_\sI^\dagger}=-\lfr_\sI$} if $\bigl(\lfr_\sI\bigr)$
is orthonormal as we'll assume.
The following table gives, for each sector,
the notation used for the related field components,
its grade (or \emph{parity}),
and the classical configuration bundles:
\begin{center}\begin{tabular}{|l|c|l|l|l|} \hline
field & grade & components & cl.\ config.\ bundle \\
\hline\hline
fermion & 1 & $\psi^{\a i}$ & $\F\tn\W \vphantom{\Bigr|}$ 
\\ \hline
gauge boson & 0 &
$A_a^\sI$ & $\TS\M\tn\Lie \vphantom{\Bigr|}$
\\ \hline
ghost & 1 & $\gh_\sI$ & $\Lie \vphantom{\Bigr|}$
\\ \hline
anti-ghost & 1 & $\agh_\sI$ & $\Lie^* \vphantom{\Bigr|}$
\\ \hline
Nakanishi-Lautrup & 0 & $n^\sI$ & $\Lie \vphantom{\Bigr|}$
\\ \hline
\end{tabular}\end{center}

\noindent
Note that the ghost, anti-ghost and Nakanishi-Lautrup (NL) fields
are all mutually independent,
the natural isomorphism \hbox{$\Lie\leftrightarrow\Lie^*$} not withstanding.

\subsection{The Lagrangian}
\label{ss:The Lagrangian}

We consider a first-order Lagrangian density
\hbox{$\Lcal=\ell\,\dO^4x:\JO\EC\to\OC\tn\weu{m}\TS\M$},
with
$$\ell=\ell_\psi+\ell_A+\ell\ghost:\JO\EC\to\OC$$
sum of fermion, gauge and ghost sectors terms defined as
\begin{align*}
&\ell_\psi=\Bigl(
\ih\,\bigl(\bar\psi_{\a i}\,\nasl\psi^{\a i}
-\nasl\bar\psi_{\a i}\,\psi^{\a i}\bigr)
-m\,\bar\psi_{\a i}\,\psi^{\a i}\Bigr) \rdg~.
\\[6pt]
&\ell_A=-\oq\,g^{ab}\,g^{cd}\,F\iI{ac}\sI\,F_{bd\sI}\,\rdg
\equiv -\oq\,g^{ab}\,g^{cd}\,F\iIi{ac}ij\,F\iIi{bd}ji\,\rdg~,
\\[6pt]
&\ell\ghost=
g^{ab}\,\agh_{\sI,a}\,\,\na_b\gh^\sI\,\rdg
+n_\sI\,(f^\sI+\oh\,\xi\,n^\sI\,)\,\rdg~,\quad \xi\in\RR~,
\end{align*}
where $g$ is the spacetime metric,
\hbox{${{\scriptstyle|}g{\scriptstyle|}}\equiv|\!\det g|$},
\hbox{$\bar\psi:\M\to\OC\tn\F^*\tn\W^*$} is the Dirac adjoint of $\psi$\,,
and
\begin{align*}
&\nasl\psi^{\a i}=
g^{ab}\,\g\iIi b\a\b\,\bigl( \psi\Ii{\b i}{,a}
-A\iIi aij\,\psi^{\b j}-\Cs\iIi a\b\g\,\psi^{\g i}\bigr)~,
\\[6pt]&
\nasl\bar\psi_{\a i}=g^{ab}\,\bigl(\bar\psi_{\b i,a}
+A\iIi aji\,\bar\psi_{\b j}+\Cs\iIi a\g\b\,\bar\psi_{\g i}\bigr)\,\g\iIi b\b\a~,
\displaybreak[2]\\[6pt]
&\Cs\iIi a\a\b=\oq\,\G\iI a{\l\m}\,(\g_\l\,\g_\m)\Ii\a\b=
\oq\,\G\iI a{\l\m}\,(\g_\l\we\g_\m)\Ii\a\b~,
\\[6pt]
&F\iI{ac}\sI=F\iIi{ac}ij\,\lfr\IiI\sI ji=
A^\sI_{[a,c]}+\sco{\sI}{\sJ}{\sH}\,A_a^\sJ\,A_c^\sH=
A^\sI_{[a,c]}+[A_a\,,\,A_c]^\sI~,
\\[6pt]
&\na_b\gh^\sI\equiv \gh^\sI_{,b}+\sco\sI\sJ\sH\,\gh^\sJ\,A_b^\sH~,
\quad f^\sI\equiv({*}\dO{*}A)^\sI=\tfrac1\rrdg\,\dO_a(g^{ab}\,\rdg\,A^\sI_b)~.
\end{align*}
Here \hbox{$\Cs:\W\to\JO\W$} denotes the \emph{spin connection},
describing the interaction between spin and gravitation.
As it is customary, for notation simplicity
we indicate field components and the related fiber coordinates
by the same symbols.

While here we are not directly involved with the field equations
for all the fields,
we note that for the NL field we get
\hbox{$n^\sI= -\tfrac1{\xi}f^\sI$}.
The NL-term in $\ell\ghost$ then becomes $-\frac1{2\xi}f_\sI f^\sI\rdg$,
which is recognized as the ``gauge fixing term''
usually added to the gauge field Lagrangian.

\subsection{BRST symmetry}
\label{ss:BRST symmetry}

We consider the morphism \hbox{$v:\JO\EC\to\VO\EC$}
which has the coordinate expression
\begin{align*}
&v=v^{\a i}\,\dde{}{\psi^{\a i}}+v_{\a i}\,\dde{}{\bar\psi_{\a i}}
+v^\sI_a\,\dde{}{A_a^\sI}+v^\sI\,\dde{}{\gh^\sI}+v_\sI\,\dde{}{\agh_\sI}~,
\\[6pt]
&v^{\a i}=\th\,\lfr\iIi\sI ij\,\gh^\sI\,\psi^{\a j}\,,~
v_{\a i}=\th\,\lfr\iIi \sI ji\,\bar\psi_{\a j}\,\gh^\sI\,,~
v^\sI_a=\th\,\na_a\gh^\sI\,,~
v^\sI=\oh\,\th\,\sco\sI\sJ\sH\,\gh^\sJ\,\gh^\sH\,,~
v_\sI=\th\,n_\sI\,,
\end{align*}
where \hbox{$\th\in\OC$} is such that $\grade{\th}=1$\,,
so that the components of $v$ have the same parities as their respective sectors.

If \hbox{$\Phi:\JO_k\EC\to\weu{q}\TS\M$} is a basic form then
(\sref{ss:Vertical infinitesimal symmetries and currents})
\hbox{$\d[v]\Phi:\JO_{k+1}\EC\to\weu{q}\TS\M$}.
We then define the \emph{BRST transformation} $\brstS$ by the rule
$$\th\,\brstS\Phi=\d[v]\Phi~.$$
In particular $\brstS$ acts on functions \hbox{$\EC\to\OC$} (\hbox{$k=q=0$}),
and for fiber coordinates we obtain essentially the usual formulas
by which it is usually introduced
as a ``transformation of the fields''~\cite{We96}:
\begin{align*}
&\brstS\psi^{\a i}=\lfr\iIi\sI ij\,\gh^\sI\,\psi^{\a j}~,\qquad
\brstS\bar\psi_{\a i}=\lfr\iIi\sI ji\,\gh^\sI\,\bar\psi_{\a j}~,\qquad
\brstS A_a^\sI=\na_a\gh^\sI=\gh^\sI_{,a}+\sco\sI\sJ\sH\,\gh^\sJ\,A_a^\sH~,
\\[6pt]
&\brstS\gh^\sI=\oh\,\sco\sI\sJ\sH\,\gh^\sJ\,\gh^\sH~,\qquad
\brstS\agh_\sI=n_\sI~,\qquad\brstS n_\sI=0~.
\end{align*}
It's not difficult then to check that $\brstS$ is nilpotent,
\hbox{$\brstS^2=0$}\,.
Consequently, for each exterior degree \hbox{$0\leq q\leq4$}
we have the exact sequence
\hbox{${}\dots\stackrel{{}_{\sst\mathrm S}}{\longrightarrow}\LA_{k}^q
\stackrel{{}_{\sst\mathrm S}}{\longrightarrow}\LA_{k+1}^q
\stackrel{{}_{\sst\mathrm S}}{\longrightarrow}\dots$}
(see \sref{ss:Vertical infinitesimal symmetries and currents}).

With regard to practical calculations
we note that proposition~\ref{prop:dHanddeltavcommute}
implies that $\d[v]$ and $\brstS$ commute with the derivation of fields
relatively to base coordinates.
We also note that the actions of $\d[v]$ and $\brstS$ on fiber polynomials are,
respectively, a derivation and an anti-derivation,
and that they respectively preserve and change grade.

We now discuss $\d[v]\Lcal$\,,
using the observation that $\Lcal\ghost$ is $\brstS$-exact
up to a horizontal differential.
\begin{proposition}
We have \hbox{$\d[v]\Lcal\ghost=\th\,\dH\Ncal$} where
$$\Ncal\equiv\bang{n,{*}\nabla\gh}=
g^{ab}\,\rdg\,n_\sI\,\na_b\gh^\sI\,\dO x_a~.$$
\end{proposition}\proof
We have \hbox{$\brstS f^\sI=\tfrac1\rrdg\,\dO_a(g^{ab}\,\rdg\,\na_b\gh^\sI)$}\,,
whence
\begin{align*}
&\brstS\bigl(\agh_\sI\,(f^\sI+\oh\,\xi\,n^\sI)\,\rdg\bigr)=
n_\sI\,(f^\sI+\oh\,\xi\,n^\sI)\,\rdg
-\agh_\sI\,\dO_a(g^{ab}\,\rdg\,\na_b\gh^\sI)=
\\[4pt]&\qquad
=n_\sI\,(f^\sI+\oh\,\xi\,n^\sI)\,\rdg
-\dO_a(\agh_\sI\,g^{ab}\,\rdg\,\na_b\gh^\sI)
+\agh_{\sI,a}\,g^{ab}\,\rdg\,\na_b\gh^\sI~,
\end{align*}
namely \hbox{$\Lcal\ghost=\brstS\Kcal+\dH\bang{\agh,{*}\nabla\gh}$} with
\hbox{$\Kcal\equiv\agh_\sI\,(f^\sI+\oh\,\xi\,n^\sI)\,\rdg\,\dO^4x$}\,.
Hence
$$\d[v]\Lcal\ghost=\th\,\brstS^2\Kcal+\th\,\brstS\dH\bang{\agh,{*}\nabla\gh}
=0+\th\,\dH\bang{n,{*}\nabla\gh}~,$$
where we used \hbox{$\d[v]\dH=\dH\d[v]$}
(proposition~\ref{prop:dHanddeltavcommute})
and \hbox{$\brstS\na_b\gh^\sI=\brstS^2A_b^\sI=0$}\,.\qed

The above proposition implies
\hbox{$\d[v]\Lcal=\d[v](\Lcal_\psi+\Lcal_A)+\th\,\dH\Ncal$}.
Now \hbox{$\Lcal_\psi\,{+}\,\Lcal_A$} is the standard Lagrangian density
of electrodynamics, dependent on spinor and e.m.\ fields
(independent of ghosts).
The action of $\d[v]$ on these is exactly that
of an infinitesimal gauge transformation,
represented by $\th\gh$\,.
Hence \hbox{$\Lcal_\psi\,{+}\,\Lcal_A$} is $\d[v]$-closed,
and finally we get \hbox{$\d[v]\Lcal=\th\,\dH\Ncal$}\,.

Recalling theorem~\ref{th:Noether} we now get a current
which we express as $\th\,\Jcal^a\,\dO x_a$\,, obtaining
$$\Jcal^a=
\bigl(-\iO\,\bang{\bar\psi\,\g^a\,\gh\,\psi}
+\bang{F^{ab},\na_b\gh}+g^{ab}\,(n_\sI\,\na_b\gh^\sI
-\oh\,\agh_{\sI,b}\,\sco\sI\sJ\sH\,\gh^\sJ\,\gh^\sH )\bigr)\,\rdg~.$$
We write \hbox{$\Jcal=\Jcal_{(\psi,A)}+\Jcal\!\ghost$} where
$\Jcal_{(\psi,A)}$ is constituted by the first two terms above;
then $\th\Jcal_{(\psi,A)}$
is the current related to an infinitesimal gauge transformation $\th\gh$
of the matter and gauge fields.
Actually, a calculation shows that \hbox{$\dH\Jcal_{(\psi,A)}=0$}\,.

\remark~The symmetry
\hbox{$\gh\to\eO^\a\,\gh$}\,, \hbox{$\agh\to\eO^{-\a}\,\agh$}\,,
\hbox{$\a\in\RR$}\,,
corresponds to the infinitesimal symmetry
\hbox{$v\spec{FP}=\gh^\sI\,\dde{}{\gh^\sI}-\agh^\sI\,\dde{}{\agh^\sI}$}\,.
We have \hbox{$\d[v\spec{FP}]\Lcal=0$} and get the \emph{Faddeev-Popov current}
$$\Jcal\!\spec{FP}=\Jcal\!\spec{FP}^{\:a}\,\dO x_a=
g^{ab}\,(\agh_{\sI,b}\,\gh^\sI+\agh_\sI\,\na_b\gh^\sI)\,\rdg\,\dO x_a~,$$
which fulfills
$$\brstS\Jcal\!\spec{FP}=
g^{ab}\,n_{\sI,b}\,\gh^\sI\,\rdg\,\dO x_a+\Jcal\!\ghost~.$$
Moreover we find \hbox{$[v\spec{FP}\,,\,v]=v$}\,.

\medbreak
With regard to the charge associated with an infinitesimal symmetry
(\sref{ss:Symmetry and charge}),
the case of the BRST symmetry is peculiar
because $v$ depends on the choice of $\th$\,,
which we get rid of by setting \hbox{$\d[v]\equiv\th\,\brstS$}
and defining the charge $\brstQ$ in such a way
that the charge in the general sense is actually $\th\brstQ$\,.
The argument of~\sref{ss:Symmetry and charge} applies because
\hbox{$\grade{\th\brstQ}=0$}\,,
and it can be checked that the equal-time identity
\hbox{$\Suc{\phi^i(x)\,,\,v^j[\phi](x')}=0$}
holds in each sector.
Hence
\begin{align*}
\th\,\brstS\phi^i&=\iO\,\bigl[\th\,\brstQ\,,\,\phi^i\bigr]\,\detg^{-1/2}=
\iO\,\th\,\Suc{\brstQ\,,\,\phi^i}\,\detg^{-1/2}
\\[6pt]
\Rightarrow\quad\brstS\phi^i&=\iO\,\Suc{\brstQ\,,\,\phi^i}\,\detg^{-1/2}\quad
\text{($i$ is a generic index).}
\end{align*}

Finally, we remark that the ghost Lagrangian is often presented
as an equivalent 2nd order density
$-\agh_{\sI}\,\dO_a(g^{ab}\,\na_b\gh^\sI\,\rdg)$\,,
rather than the 1st order density
$g^{ab}\,\agh_{\sI,a}\,\,\na_b\gh^\sI\,\rdg$\,.
Actually this amounts to considering a new Lagrangian
\hbox{$\Lcal'=\Lcal-\dH\Mcal$} where
$$\Mcal\equiv\bang{\agh,{*}\nabla\gh}=
g^{ab}\,\agh_{\sI}\,\na_b\gh^\sI\,\rdg\,\dO x_a~.$$
We immediately see that \hbox{$\d[v]\Mcal=\th\Ncal$}, so that
$$\d[v]\Lcal'=\d[v]\Lcal-\d[v]\dH\Mcal=\th\,\dH\Ncal-\dH\d[v]\Mcal=0~.$$
Hence (theorem~\ref{th:Noether}) we obtain the 2-nd order current
$$\Pcal^a\pint v_{\sst(1)}=(P^a_i-\dO_b P^{ab}_i)\,v^i+P^{ab}_i\,\dO_b v^i~,$$
which, by a straightforward calculation,
can be shown to coincide with $\th\,\Jcal$.

\end{document}